\pdfoutput=1
\newcommand*{\ATLASLATEXPATH}{latex/}
\documentclass[cernpreprint, texlive=2016, UKenglish]{\ATLASLATEXPATH atlasdoc}

\usepackage[backend=bibtex]{\ATLASLATEXPATH atlaspackage}
\usepackage{\ATLASLATEXPATH atlasbiblatex}

\usepackage{\ATLASLATEXPATH atlasphysics}

\graphicspath{{logos/}{figures/}}

\usepackage{WZVBS-defs}

\AtlasTitle{Observation of electroweak \wz boson pair production in association with two jets in $pp$ collisions at $\rts = 13~\tev$ with the ATLAS detector}

\AtlasAbstract{%
  An observation of electroweak \wz\ production in association with two jets in proton--proton collisions is presented.
The data collected by the ATLAS detector at the Large Hadron Collider in $2015$ and $2016$ at a centre-of-mass energy of $\rts = 13~\tev$ are used, corresponding to
an integrated luminosity of $36.1~\ifb$.
Events containing three identified leptons, either electrons or muons, and two jets are selected.
The electroweak production of \wz bosons in association with two jets is measured with an observed significance of $5.3	$ standard deviations.
A fiducial cross-section for electroweak production including interference effects and for a single leptonic decay mode is measured to be $\sigma_{WZjj\mathrm{-EW}} = 0.57 \; ^{+ 0.14} _{- 0.13} \,(\mathrm{stat.}) \; ^{+ 0.07} _{- 0.06} \,(\mathrm{syst.}) \; \mathrm{fb}$.
Total and differential fiducial cross-sections of the sum of \wzjj electroweak and strong productions for several kinematic observables are also measured.
}

\AtlasRefCode{STDM-2017-23}

\PreprintIdNumber{CERN-EP-2018-286}

\AtlasJournalRef{\PLB 793 (2019) 469}
\AtlasDOI{10.1016/j.physletb.2019.05.012}

\hypersetup{pdftitle={ATLAS document},pdfauthor={The ATLAS Collaboration}}

\begin{document}

\maketitle


\section{Introduction}
\label{sec:intro}

The scattering of vector bosons (VBS), $VV \to VV$ with $V= W/Z/\gamma$, is a key process with which to probe the $\text{SU}(2)_{\mathrm{L}} \times \text{U}(1)_{\mathrm{Y}}$ gauge symmetry of the electroweak (EW) theory  that determines the self-couplings of the vector bosons.
%
New phenomena beyond the Standard Model (SM) can alter the couplings of vector bosons, generating additional contributions to quartic gauge couplings (QGC) compared with the SM predictions~\cite{Eboli:2003nq,Eboli:2006wa,Degrande:2012wf}.

In proton--proton collisions, VBS is initiated by an interaction of two vector bosons radiated from the initial-state quarks, yielding a final state with two bosons and two jets, $VVjj$, in a purely electroweak process~\cite{Cahn:1986zv}.
VBS diagrams are not independently gauge invariant and cannot be studied separately from other processes leading to the same $VVjj$ final state~\cite{Accomando:2006mc}.
Two categories of processes give rise to $VVjj$ final states. The first category, which includes VBS contributions, involves exclusively weak interactions at Born level of order $\alpha_{\mathrm{EW}}^6$ including the boson decays, where $\alpha_{\mathrm{EW}}$ is the electroweak coupling constant.
It is referred to as electroweak production.
The second category involves both the strong and electroweak interactions at Born level of order $\alphas^2\alpha_{\mathrm{EW}}^4$, where $\alphas$ is the strong interaction coupling constant.
It is referred to as QCD production.
According to the SM a small interference occurs between electroweak and QCD production.

Different searches for diboson electroweak production have been performed by the ATLAS and CMS collaborations at the LHC. 
So far, electroweak $VVjj$ production has only been observed in the same-sign $W^\pm W^\pm jj$ channel by CMS using data collected at a centre-of-mass energy of $\sqrt{s} = 13$~\TeV~\cite{Sirunyan:2017ret}.
Evidence of electroweak $VVjj$ production has also been obtained in the $W^\pm W^\pm jj $~\cite{STDM-2013-06,Aaboud:2016ffv} and $Z\gamma jj$~\cite{Khachatryan:2017jub} channels by ATLAS and CMS, respectively, using smaller samples of data recorded at  $\sqrt{s} = 8$~\TeV.
Limits on electroweak cross-sections for the production of two gauge boson have been reported for the \wzjj~\cite{ATLASWZ_8TeV,Sirunyan:2019ksz}, $ZZjj$~\cite{CMS-SMP-17-006}, $Z\gamma jj$~\cite{STDM-2015-21} and $W\gamma jj$~\cite{Khachatryan:2016vif} channels by ATLAS and CMS.

This Letter reports on an observation and measurement of electroweak \wzjj production, exploiting the fully leptonic final states where both the $Z$ and $W$ bosons decay into electrons or muons.
The $pp$ collision data were collected with the ATLAS detector in $2015$ and $2016$ at a centre-of-mass energy of $\sqrt{s} = 13$~\TeV\ and correspond to an integrated luminosity of $36.1~\ifb$.

\section{The ATLAS detector}
\label{sec:detector}

The ATLAS detector~\cite{PERF-2007-01}  is a multipurpose detector with a cylindrical geometry\footnote{ATLAS uses a right-handed coordinate system with its origin at the nominal 
interaction point (IP) in the centre of the detector and the $z$-axis along the beam direction. 
The $x$-axis points from the IP to the centre of the LHC ring, and the $y$-axis points upward. 
Cylindrical coordinates $(r,\phi)$ are used in the transverse $(x,y)$ plane, $\phi$ being the 
azimuthal angle around the beam direction. The pseudorapidity is defined in terms of the
 polar angle $\theta$ as $\eta = -\textrm{ln}[ \textrm{tan}(\theta/2)]$. 
} and nearly $4\pi$ coverage in solid 
angle.
The collision point is surrounded by inner tracking detectors, collectively referred to as the inner detector (ID), located within a superconducting solenoid providing a $2$~T axial magnetic field, followed by a calorimeter system and a muon spectrometer (MS).

The inner detector provides precise measurements of charged-particle tracks in the pseudorapidity range $|\eta| < 2.5$.
It consists of three subdetectors arranged in a coaxial geometry around the beam axis: a silicon pixel detector, %
a silicon microstrip detector and a transition radiation tracker. 

The electromagnetic calorimeter covers the region $|\eta|<3.2$ and is based on high-granularity, lead/liquid-argon (LAr) sampling technology. 
The hadronic calorimeter uses a steel/scintillator-tile detector in the region $|\eta|<1.7$ and a copper/LAr detector in the region $1.5 < |\eta| < 3.2$. 
The most forward region of the detector, $3.1 <|\eta| < 4.9$, is equipped with 
a forward calorimeter, measuring electromagnetic and hadronic energies in copper/LAr and tungsten/LAr modules.

The muon spectrometer comprises separate trigger 
and high-precision tracking chambers to measure the deflection of muons in a magnetic field generated by three 
large superconducting toroidal magnets arranged with an eightfold azimuthal coil symmetry around the calorimeters. 
The high-precision chambers cover the range $|\eta|< 2.7$  with three layers of monitored drift tubes, complemented 
by cathode strip chambers in the forward region, where the particle flux is highest.
The muon trigger system covers the range $|\eta|< 2.4$ with 
resistive-plate chambers in the barrel and thin-gap chambers in the endcap regions.

A two-level trigger system is used to select events in real time~\cite{Aaboud:2016leb}. It consists of a hardware-based first-level trigger and a software-based high-level trigger. The latter employs algorithms similar to those used offline to identify electrons, muons, photons and jets.


\section{Phase space for cross-section measurements}
\label{sec:phase_space}

The \wzjj electroweak cross-section is measured in a fiducial phase space
that is defined by the kinematics of the final-state leptons, electrons or muons, associated with the \Wpm and $Z$ boson decays, and of two jets.
Leptons produced in the decay of a hadron, a $\tau$-lepton, or their descendants are not considered in the definition of the fiducial phase space.
At particle level, the kinematics of the charged lepton after quantum electrodynamics (QED) final-state radiation (FSR) are `dressed'
by including contributions from photons with an angular distance $\Delta R \equiv \sqrt{(\Delta\eta)^2 + (\Delta\phi)^2} < 0.1$
from the lepton. Dressed charged leptons, and final-state neutrinos that do not originate from hadron or $\tau$-lepton decays,
are matched to the \Wpm and $Z$ boson decay products using a Monte Carlo (MC) generator-independent algorithmic approach,
called the `resonant shape' algorithm. This algorithm is based on the value of an estimator expressing the product of the nominal line shapes of the $W$ and $Z$ resonances as detailed in Ref.~\cite{ATLASWZ_8TeV}.

The fiducial phase space of the measurement matches the one used in Refs.~\cite{ATLASWZ_8TeV,STDM-2015-19} and is
defined at particle level by the following requirements:
the charged leptons from the $Z$ boson decay are required to have transverse momentum $\pt>15$~\GeV, the charged lepton from the \Wpm decay is required to have transverse momentum $\pt^\ell>20$~\GeV,
the charged leptons from the \Wpm and $Z$ bosons are required to have $\abseta<2.5$ and
the invariant mass of the two leptons from the $Z$ boson decay must be within $10$~\GeV{}
of the nominal Z boson mass, taken from the world average value, $m_Z^{\textrm{PDG}}$~\cite{Tanabashi:2018oca}.
The $W$ boson transverse mass, defined as $m_{\mathrm{T}}^{W} = \sqrt{2 \cdot p_{\mathrm{T}}^\nu \cdot p_{\mathrm{T}}^\ell \cdot [1 -\cos{\Delta \phi(\ell, \nu)}]}$, where $\Delta \phi(\ell, \nu)$
is the angle between the lepton and the neutrino in the transverse plane and $\pt^\nu$ is the transverse momentum of the neutrino, is required to be $m_{\mathrm{T}}^{W}>30$~\GeV.
The angular distance between the charged lepton from the $W^\pm$ decay and each of the charged leptons from the $Z$ decay is
required to be  $\Delta R>0.3$, and the angular distance between the two leptons from the $Z$ decay is required to be $\Delta R>0.2$.
Requiring that the transverse momentum of the leading lepton be above  $27$~\GeV\ reduces the acceptance of the fiducial phase space by only $0.02\%$ and is therefore not added to the definition of the fiducial phase space, although it is present in the selection at the detector level presented in Section~\ref{sec:selection}.

In addition to these requirements that define an inclusive phase space, at least two jets with $\pt>40$~\GeV\ and $|\eta_j|< 4.5$ are required.
These particle-level jets are reconstructed from stable particles
with a lifetime of $\tau>30$~ps in the simulation after parton showering, hadronisation, and decay of particles with $\tau<30$~ps.
Muons, electrons, neutrinos and photons associated with $W$ and $Z$ decays are excluded.
The particle-level jets are reconstructed using the anti-$k_{t}$~\cite{antikt} algorithm with a radius parameter $R$ = $0.4$.
The angular distance between all selected leptons and jets is required to be $\Delta R (j,\ell)>0.3$.
If the $\Delta R (j,\ell)$ requirement is not satisfied, the jet is discarded.
The invariant mass, $m_{jj}$, of the two highest-\pT jets in opposite hemispheres, $\eta_{j1} \cdot \eta_{j2} < 0$, is required to be $m_{jj}>500$~\GeV\ to enhance the sensitivity to VBS processes.
These two jets are referred to as tagging jets.
Finally, processes with a $b$-quark in the initial state, such as $tZj$ production,  are not considered as signal.
The production of $tZj$ results from a $t$-channel exchange of a $W$ boson  between a $b$- and a $u$-quark giving a final state with a $t$-quark, a $Z$ boson and a light-quark jet, but does not include diagrams with gauge boson couplings.
%

\section{Signal and background simulation}
\label{sec:samples}

Monte Carlo simulation is used to model signal and background processes.
All generated MC events were passed through the ATLAS detector simulation~\cite{Aad:2010ah}, based on
\GEANT4~\cite{Agostinelli:2002hh}, and processed using the same reconstruction software as used for the data.
The event samples include the simulation of additional proton--proton interactions (pile-up)
generated with \PYTHIA~$8.186$~\cite{Sjostrand:2007gs} using the \texttt{MSTW2008LO}~\cite{Martin:2009iq} parton distribution functions (PDF)
and the A2~\cite{ATL-PHYS-PUB-2012-003} set of tuned parameters. %
Scale factors are applied to simulated events to correct for the differences between data and MC simulation
in the trigger, reconstruction, identification, isolation and impact parameter efficiencies of electrons
and muons~\cite{ATLAS-CONF-2016-024,PERF-2015-10}.
Furthermore, the electron energy and muon momentum in simulated events are smeared to account for
differences in resolution between data and MC simulation~\cite{PERF-2015-10,Aaboud:2018ugz}.   %

The \Sherpa~$2.2.2$   MC event generator~\cite{Gleisberg:2008ta, Gleisberg:2008fv, Schumann:2007mg, Catani:2001cc, Hoeche:2009rj, Hoeche:2012yf, Winter:2003tt, Schonherr:2008av} was used to model \wzjj\ events.
It includes the modelling of hard scattering, parton showering, hadronisation and the underlying event.
A MC event sample, referred to as \wzjjew, includes processes of order six (zero) in $\alpha_\mathrm{EW}$ ($\alphas$).
In this sample, which includes  VBS diagrams, two additional jets originating from electroweak vertices from matrix-element partons are included in
the final state.
Diagrams with a $b$-quark in either the initial or final state, i.e.\ $b$-quarks in the matrix-element calculation, are not considered.
This sample provides a \LO prediction for the \wzjjew signal process.
A second MC event sample, referred to as \wzjjqcd, includes processes of order four in $\alpha_\mathrm{EW}$ in the matrix-element of \wz production with up to one jet calculated at next-to-leading order (\NLO) and with a second or third jet calculated at leading order (\LO).
This \wzjjqcd sample includes matrix-element $b$-quarks.
Both \sherpa samples were generated using the \texttt{NNPDF3.0}~\cite{Ball:2014uwa} PDF set.
Interferences between the two processes were estimated at \LO using the  \MGMCatNLO~$2.3$~\cite{Alwall:2014hca} MC event generator with the \texttt{NNPDF3.0} PDF set, including only contributions to the squared matrix-element of order one in $\alphas$.
They are found to be positive and approximately $10\%$ of the \wzjjew cross-section in the fiducial phase space and are treated as an uncertainty in the measurement, as discussed in Section~\ref{sec:systematics}.
For the estimation of modelling uncertainties, alternative MC samples of \wzjjqcd and \wzjjew processes were generated with \MGMCatNLO~$2.3$ at LO in QCD, including up to two partons in the matrix-element for \wzjjqcd, and using the \texttt{NNPDF3.0} PDF set.
MC samples of inclusive \wz production generated at NLO in QCD with the \POWHEG-\textsc{Box}~v2~\cite{powheg1:2004, powheg, Alioli:2010xd, Melia:2011tj} generator, interfaced to \pythia~8.210 or \herwigpp~2.7.1~\cite{Bahr:2008pv} for simulation of parton showering and hadronisation are also used for tests of the modelling of \wzjjqcd events.

The $q\bar{q} \to ZZ^{(\ast)}$ processes were generated with \SHERPA~$2.2.2$ and the \texttt{NNPDF3.0} PDF set.
Similarly to \wz\ simulation, the \zzjjqcd\ and \zzjjew\ processes are generated separately with the same matrix-element accuracy as for the \wz \sherpa MC samples.
The \SHERPA~$2.1.1$ MC event generator was used to model the $gg\to ZZ^{(\ast)}$ 
and $VVV$ processes at \LO using the \texttt{CT10}~\cite{CT10} PDF set.
The $t\bar{t}V$ processes were generated at NLO with the \MGMCatNLO~$2.3$ MC generator using the \texttt{NNPDF3.0} PDF set interfaced to the \PYTHIA~8.186 parton shower model.
  The associated production of a single top quark and a $Z$ boson was simulated at LO with \MGMCatNLO~$2.3$ using the \texttt{NNPDF3.0} PDF set and interfaced to \PYTHIA~$8.186$ for parton shower.
%

\section{Event selection}
\label{sec:selection}

Candidate events were selected using single-leptons triggers~\cite{Aaboud:2016leb} that required at least one electron or muon.
The transverse momentum threshold of the leptons
in $2015$ was $24$~\GeV\ for electrons and $20$~\GeV\ for muons satisfying a loose isolation requirement based only on ID track information.
Due to the higher instantaneous luminosity in $2016$ the trigger threshold was increased to $26$~\GeV\ for
both the electrons and muons and tighter isolation requirements were applied.
Possible inefficiencies for leptons with large transverse momenta were reduced by including additional electron and muon
triggers that did not include any isolation requirements with transverse momentum thresholds of $\pT = 60$~\GeV\ and $50$~\GeV, respectively.
Finally, a single-electron trigger requiring $\pT>120$~\GeV\ or $\pT>140$~\GeV\ in $2015$ and $2016$, respectively, with less restrictive electron identification criteria was used to increase the selection efficiency for high-$\pt$ electrons.
The combined efficiency of these triggers is close to $100\%$ for \wzjj events.
Only data recorded with stable beam conditions and with all relevant detector subsystems operational are considered.

Events are required to have a primary vertex reconstructed from at least two charged-particle tracks and compatible with the \pp\ interaction region.
If several such vertices are present in the event, the one with the highest sum of the $\pT^{2}$ of the associated tracks is selected as the production vertex of the \wz.
All final states with three charged leptons (electrons or muons) and neutrinos from \wz\ leptonic decays are considered.
%


Muon candidates are identified by tracks reconstructed in the muon spectrometer and matched to tracks reconstructed in the inner detector. Muons are required to satisfy a `medium' identification selection that is based on requirements on the number of hits in the ID and the MS~\cite{PERF-2015-10}. The efficiency of this selection averaged over \pt and $\eta$ is $>98\%$.
The muon momentum is calculated by combining the MS measurement, corrected for the energy deposited in the calorimeters, with the ID measurement.
The transverse momentum of the muon must satisfy $\pt>15$~\GeV\ and its pseudorapidity must satisfy $|\eta|<2.5$.

Electron candidates are reconstructed from energy clusters in the electromagnetic calorimeter matched to ID tracks. Electrons are identified using a likelihood function constructed from information from the shape of the electromagnetic showers in the calorimeter, track properties and track-to-cluster matching quantities~\cite{ATLAS-CONF-2016-024}. Electrons must satisfy a `medium' likelihood requirement, which provides an overall identification efficiency of $90\%$.
The electron momentum is computed from the cluster energy and the direction of the track. The transverse momentum of the electron must satisfy $\pt>15$~\GeV\ and the pseudorapidity of the cluster must be in the ranges $\abseta< 1.37$ or $1.52 < \abseta < 2.47$.

Electron and muon candidates are required to originate from the primary vertex. The significance of the track's transverse impact parameter relative to the beam line must satisfy $|{d_{0}/\sigma_{d_{0}}}| <3\ (5)$ for muons (electrons), and the longitudinal impact parameter, $z_{0}$ (the difference between the value of $z$ of the point on the track at which $d_{0}$ is defined and the longitudinal position of the primary vertex), is required to satisfy $|z_0\cdot \sin(\theta)|<0.5$~mm.
Electrons and muons are required to be isolated from other particles, according to calorimeter-cluster and ID-track information.
The isolation requirement for electrons varies with \pt\ and is tuned for an efficiency of  at least $90\%$ for $\pT>25$~\GeV\ and
at least $99\%$ for $\pT>60$~\GeV~\cite{ATLAS-CONF-2016-024}. Fixed thresholds values are used for the muon isolation variables,  providing an efficiency above $90\%$ for $\pT>15$~\GeV\ and at least $99\%$ for $\pT>60$~\GeV~\cite{PERF-2015-10}.

Jets are reconstructed from clusters of energy depositions in the calorimeter~\cite{PERF-2014-07} using the anti-$k_{t}$ algorithm~\cite{antikt} with a radius parameter $R=0.4$.
Events with jets arising from detector noise or other non-collision sources are discarded~\cite{ATLAS-CONF-2015-029}.
All jets must have $\pt>25$~\GeV\ and be reconstructed in the pseudorapidity range $|\eta|<4.5$.
A multivariate combination of track-based variables is used to suppress jets originating from pile-up in the ID acceptance~\cite{Aad:2015ina}.
The energy of jets is calibrated using a jet energy correction derived from simulation
and {\textit{in situ}} methods using data~\cite{Aaboud:2017jcu}. %
Jets in the ID acceptance with $\pt>25$~\GeV\  containing a $b$-hadron are identified using a multivariate algorithm~\cite{PERF-2012-04,ATL-PHYS-PUB-2016-012}  that uses impact parameter and reconstructed secondary vertex information of the tracks contained in the jets.
Jets initiated by $b$-quarks are selected by setting the algorithm's output threshold such that a $70\%$ $b$-jet selection efficiency is achieved in simulated $t\bar{t}$ events.
%
%

The transverse momentum of the neutrino is estimated from the missing transverse momentum in the event, \met,
calculated as the negative vector sum of the transverse momentum of all identified hard (high \pt) physics objects
(electrons, muons and jets), as well as an additional soft term. A track-based measurement of the soft term~\cite{Aaboud:2018tkc,ATLAS-CONF-2018-023}, %
which accounts for low-$\pt$ tracks not assigned to a hard object, is used.
%
%


Events are required to contain exactly three lepton candidates satisfying the selection criteria described above.
To ensure that the trigger efficiency is well determined, at least one of the candidate leptons is required to have
$\pt > 25$~\GeV\ or $\pt > 27$~\GeV\ for the $2015$ or $2016$ data, respectively, and to be geometrically matched to a lepton that was selected by the trigger.

To suppress background processes with at least four prompt leptons, events with a fourth lepton candidate
satisfying looser selection criteria are rejected. For this looser selection, the $\pt$ requirement for the leptons is
lowered to $\pt>5$~\GeV{} and `loose' identification requirements are used for both the electrons and muons.
A less stringent requirement is applied for electron isolation based only on ID track information and electrons with cluster in the range  $1.37 \le \abseta \le 1.52$ are also considered.

Candidate events are required to have at least one pair of leptons of the same flavour and of opposite charge, with an invariant mass that is consistent with the nominal \Zboson\ boson mass~\cite{Agashe:2014kda} to within $10$~\gev. This pair is considered to be the $Z$ boson candidate. If more than one pair can be formed, the pair whose invariant mass is closest to the nominal $Z$ boson mass is taken as the \Zboson\ boson candidate.

The remaining third lepton is assigned to the $W$ boson decay. The transverse mass of the $W$ candidate, computed using \met and the \pt of the associated lepton, is required to be greater than $30$~\GeV.

Backgrounds originating from misidentified leptons are suppressed by requiring the lepton associated with the $W$ boson to satisfy more stringent selection criteria. Thus, the transverse momentum of these leptons is required to be $\pt>20$~\GeV.
Furthermore, leptons associated with the $W$ boson decay are required to satisfy the `tight'
identification requirements, which have an efficiency between $90\%$ and $98\%$ for muons and
an efficiency of $85\%$ for electrons. %
Finally, muons must also satisfy a tighter isolation requirement, tuned for an efficiency of at least $90\%$~($99\%$) for $\pT>25~(60)$~\GeV.


To select \wzjj\ candidates, events are further required to be associated with at least two `tagging' jets.
The leading tagging jet is selected as the highest-\pt\ jet in the event with $\pt > 40$~\GeV.
The second tagging jet is selected as the one with the highest \pt\ among the remaining jets that have a pseudorapidity of opposite sign to the first tagging jet and a $\pt > 40$~\GeV.
These two jets are required to verify $m_{jj} > 150$~\GeV, in order to minimise the contamination from triboson processes.

The final signal region (SR) for VBS processes is defined by requiring that the invariant mass of the two tagging jets, $m_{jj}$, be above $500$~\GeV\ and that no $b$-tagged jet be present in the event.

\section{Background estimation}
\label{sec:background}

The background sources are classified into two groups: events where at least one of the candidate leptons is not a prompt lepton (reducible background) and events where all candidates are prompt leptons or are produced in the decay of a $\tau$-lepton (irreducible background). Candidates that are not prompt leptons are also called  `misidentified' or `fake' leptons.

The dominant source of background originates from the QCD-induced production of \wz dibosons in association with two jets, \wzjjqcd.
The shapes of distributions of kinematic observables of this irreducible background are modelled by the \sherpa MC simulation.
The normalisation of this background is, however, constrained by data in a dedicated control region.
This region, referred to as \wzjjqcd CR, is defined by selecting a sub-sample of \wzjj candidate events with  $m_{jj} < 500$~\GeV\ and no reconstructed $b$-jets.

The other main sources of irreducible background arise from $ZZ$ and $t\bar{t} + V$ (where $V$ = $Z$ or $W$).%
These irreducible backgrounds are also modelled using MC simulations.
 Data in two additional dedicated control regions, referred to as $ZZ$-CR and $b$-CR, respectively, are used to constrain the normalisations of the \zzjjqcd\ and $t\bar{t} + V$ backgrounds.
The control region $ZZ$-CR, enriched in $ZZ$ events, is defined by applying the \wzjj event selection defined in Section~\ref{sec:selection}, with the exception that instead of vetoing a fourth lepton it is required that events have at least a fourth lepton candidate with looser identification requirements.
This region is dominated by \zzjjqcd events with a small contribution of \zzjjew\ events.
The control region $b$-CR, enriched in $t\bar{t} + V$ events, is defined by selecting \wzjj\ candidate events having at least one reconstructed $b$-jet. 
Remaining sources of irreducible background are \zzjjew\, $VVV$ and $t Zj$ events. 
Their contributions in the control and signal regions are estimated from MC simulations.

The reducible backgrounds originate from $Z+j$, $Z\gamma$, $t\bar{t}$,
$Wt$ and $WW$ production processes.
The reducible backgrounds are estimated using a data-driven method based on the inversion of a global matrix containing the
 efficiencies and the misidentification probabilities for prompt and fake leptons~\cite{ATLASWZ_8TeV,Aad:2014pda}.
The method exploits  the classification of the lepton as  loose  or tight  candidates
and  the probability that a fake lepton is misidentified as a loose or tight
lepton candidate.
Tight leptons candidates are signal lepton candidates as defined in Section~\ref{sec:selection}.
Loose lepton candidates are leptons that do not meet the isolation and identification criteria of signal lepton candidates but satisfy only looser criteria.
The misidentification probabilities for fake leptons are determined from data, using dedicated control samples enriched in non-prompt leptons from heavy-flavour jets and in misidentified leptons from photon conversions or charged hadrons in light-flavour jets.
The lepton misidentification probabilities are applied to samples of \wzjj candidate events in data where at least one and up to three of the lepton candidates are loose.
Then, using a matrix inversion, the number of events with at least one misidentified lepton, which represents
 the amount of reducible background in the selected \wzjj sample, is obtained.

The number of observed events together with the expected background contributions are summarised in Table~\ref{tab:VBSyield_prefit} for the signal region and the three control regions.
All sources of uncertainties, as described in Section~\ref{sec:systematics}, are included.
The expected signal purity in the \wzjj signal region is about $13\%$, and $72\%$ of the events arise from \wzjjqcd production.

\begin{table}[!htbp] 
\caption{Expected and observed numbers of events in the \wzjj signal region and in the three control regions, before the fit.
The expected number of \wzjjew events from \sherpa and the estimated number of background events from the other processes are shown.
The sum of the backgrounds containing misidentified leptons is labelled `Misid. leptons'.
The contribution arising from interferences between \wzjjqcd and \wzjjew processes is not included in the table.
The total uncertainties are quoted.
\label{tab:VBSyield_prefit}
}
\centering
\begin{tabular}{%
l
S@{$\pm$}S
S@{$\pm$}S
S@{$\pm$}S
S@{$\pm$}S
}
\toprule
                      &  \multicolumn{2}{c}{SR}	  &  \multicolumn{2}{c}{\wzjjqcd CR}	 &  \multicolumn{2}{c}{$b$-CR} & \multicolumn{2}{c}{$ZZ$-CR}\\
\midrule
Data                  & \multicolumn{2}{c}{161}   & \multicolumn{2}{c}{213}              & \multicolumn{2}{c}{141}     & \multicolumn{2}{c}{52}     \\
Total predicted       &  200   & 41               &  290   &  61                         &  160 & 14                   &  45.2 & 7.5                \\
\midrule                                                                                                                                            
\wzjjew  (signal)     &  24.9  & 1.4              &  8.45  & 0.37                        &  1.36  & 0.10               &  0.21 & 0.12               \\
\wzjjqcd              &  144   & 41               &  231   & 60                          &  24.4  & 1.7                &  1.43 & 0.22               \\
Misid. leptons        &  9.8   & 3.9              &  17.7  & 7.1                         &  30  & 12                   &  0.47 & 0.21               \\
\zzjjqcd	      &  8.1   & 2.2              &  15.0  & 3.9                         &  1.96  & 0.49               &  35   & 11                 \\
$tZj$	              &  6.5   & 1.2              &  6.6   & 1.1                         &  36.2  & 5.7                &  0.18 & 0.04               \\
$t\bar{t}+V$          &  4.21  & 0.76             &  9.11  & 1.40                        &  65.4  & 10.3               &  2.8  & 0.61               \\
\zzjjew 	      &  1.80  & 0.45             &  0.53  & 0.14                        &  0.12  & 0.09               &  4.1  & 1.4                \\
$VVV$	  	      &  0.59  & 0.15             &  0.93  & 0.23                        &  0.13  & 0.03               &  1.05 & 0.30               \\
\bottomrule
\end{tabular}
\end{table}

\section{Signal extraction procedure}
\label{sec:analysis}

%
%
Given the small contribution to the signal region of \wzjjew processes, a multivariate discriminant is used to separate the signal from the backgrounds.
A boosted decision tree (BDT), as implemented in the TMVA package~\cite{Hocker:2007ht}, is used to exploit the kinematic differences between the \wzjjew signal and the \wzjjqcd and other backgrounds.
The BDT is trained and optimised on simulated events to separate \wzjjew events from all background processes.

A total of $15$ variables are combined into one discriminant, the BDT score output value in the range $[-1,1]$. %
The variables can be classified into three categories: jet-kinematic variables, vector-bosons-kinematics variables, and variables related to both jets and leptons kinematics.
The variables related to the kinematic properties of the two tagging jets are the invariant mass of the two jets, $m_{jj}$, the transverse momenta of the jets, the difference in pseudorapidity and azimuthal angle between the two jets, $\Delta \eta_{jj}$ and $\Delta \phi_{jj}$, the rapidity of the leading jet and the jet multiplicity.
Variables related to the kinematic properties of the vector bosons are the transverse momenta of the $W$ and $Z$ bosons, the pseudorapidity of the $W$ boson, the absolute difference between the rapidities of the $Z$ boson and the lepton from the decay of the $W$ boson, $|y_Z - y_{\ell,W}|$, and the transverse mass of the \wz system \mtwz.
The pseudorapidity of the $W$ boson is reconstructed using an estimate of the longitudinal momentum of the neutrino obtained using the $W$ mass constraint as detailed in Ref.~\cite{Aad:2012twa}.
The \mtwz observable is reconstructed following Ref.~\cite{ATLASWZ_8TeV}.
Variables that relate the kinematic properties of jets and leptons are the distance in the pseudorapidity--azimuth plane between the $Z$ boson and the leading jet, $\Delta R(j_1,Z)$,
the event balance $R_{p_\mathrm{T}}^{\mathrm{hard}}$, defined as the transverse component of the vector sum of the $WZ$ bosons and tagging jets momenta, normalised to their scalar \pt sum, and, finally the centrality of the $WZ$ system relative to the tagging jets, defined as $\zeta_{\mathrm{lep.}} = \text{min}(\Delta\eta_{-},\Delta\eta_{+})$, with $\Delta\eta_{-} = \text{min}(\eta_{\ell}^{W},\eta_{\ell_2}^{Z},\eta_{\ell_1}^{Z}) - \text{min}(\eta_{j_1},\eta_{j_2})$ and $\Delta\eta_{+} = \text{max}(\eta_{j_1},\eta_{j_2}) - \text{max}(\eta_{\ell}^{W},\eta_{\ell_2}^{Z},\eta_{\ell_1}^{Z})$.
A larger set of discriminating observables was studied but only variables improving signal-to-background were retained.
The good modelling by MC simulations of the distribution shapes and the correlations of all input variables to the BDT is verified in the \wzjjqcd CR, as exemplified by the good description of the BDT score distribution of data in the \wzjjqcd CR shown in Figure~\ref{fig:BDTScore_WZjjQCDCR}.

The distribution of the BDT score in the \wzjj signal region is used to extract the significance of the \wzjjew signal and to measure its fiducial cross-section via a maximum-likelihood fit.
An extended likelihood is built from the product of four likelihoods corresponding to the BDT score distribution in the \wzjj SR, the $m_{jj}$ distribution in the \wzjjqcd CR, the multiplicity of reconstructed $b$-quarks in the $b$-CR and the $m_{jj}$ distribution in the $ZZ$-CR.
The inclusion of the three control regions in the fit allows the yields of the \wzjjqcd, $t\bar{t}+V $ and \zzjjqcd backgrounds to be constrained by data.
The shapes of these backgrounds are taken from MC predictions and can vary within the uncertainties affecting the measurement as described in Section~\ref{sec:systematics}.
The normalisations of these backgrounds are introduced in the likelihood as parameters, labelled $\mu_{\wzjjqcd}$, $\mu_{t\bar{t}+V}$ and $\mu_{\zzjjqcd}$ for
\wzjjqcd, $t\bar{t}+V $ and \zzjjqcd backgrounds, respectively.
They are treated as unconstrained nuisance parameters that are determined mainly by the data in the respective control region.
The normalisation and shape of the other irreducible backgrounds are taken from MC simulations and are allowed to vary within their respective uncertainties.
The distribution of the reducible background is estimated from data using the matrix method presented in Section~\ref{sec:background} and is allowed to vary within its uncertainty.

The determination of the fiducial cross-section is carried out using the signal strength parameter $\mu_{\wzjjew}$:

\[
\mu_{\wzjjew} = \frac{ N^{\mathrm{signal}}_{\mathrm{data}} } { N^{\mathrm{signal}}_{\mathrm{MC}} } = \frac{ \sigma_{\wzjjew}^{\mathrm{fid.}} } { \sigma_{\wzjjew}^{\mathrm{fid., MC}} } \; ,
\label{eq:muEW}
\]
where $N^{\mathrm{signal}}_{\mathrm{data}}$ is the signal yield extracted from data by the fit and $N^{\mathrm{signal}}_{\mathrm{MC}}$ is the number of signal events predicted by the \sherpa MC simulation.
The measured cross-section $\sigma_{\wzjjew}^{\mathrm{fid.}}$ is derived from the signal strength $\mu_{\wzjjew}$ by multiplying it by the \sherpa MC cross-section prediction $\sigma_{\wzjjew}^{\mathrm{fid.,\, MC}}$ in the fiducial region.
The \wzjjqcd contribution that is considered as background in the fit procedure does not contain interference between the \wzjjqcd and \wzjjew processes.
The measured cross-section $\sigma_{\wzjjew}^{\mathrm{fid.}}$ therefore formally corresponds to the cross-section of the electroweak production including interference effects.

\section{Systematic uncertainties}
\label{sec:systematics}

Systematic uncertainties in the signal and control regions affecting the shape and normalisation of the BDT score, $m_{jj}$ and $N_{b\textrm{-jets}}$ distributions for the individual backgrounds, as well as the acceptance of the signal and the shape of its template are considered.
If the variation of a systematic uncertainty as a function of the BDT score is consistent with being due to statistical fluctuations, this systematic uncertainty is neglected.

Systematic uncertainties due to the theoretical modelling in the event generator used to evaluate the \wzjjqcd and \wzjjew templates are considered.
Uncertainties due to higher order QCD corrections are evaluated by varying the renormalisation and factorisation scales independently by factors of two and one-half, removing combinations where the variations differ by a factor of four.
These uncertainties are of $-20\%$ to $+30\%$ on the \wzjjqcd background normalisation and up to $\pm 5\%$ on the \wzjjew signal shape.
The uncertainties due to the PDF and the $\alphas$ value used in the PDF determination are evaluated using the \texttt{PDF4LHC} prescription~\cite{Butterworth:2015oua}. They are of the order of $1\%$ to $2\%$ in shape of the predicted cross-section.
A global modelling uncertainty in the \wzjjqcd background template that includes effects of the parton shower model is estimated by comparing predictions of the BDT score distribution in the signal region from the \sherpa and \madgraph MC event generators.
The difference between the predicted shapes of the BDT score distribution from the two generators is considered as an uncertainty. 
The resulting uncertainty ranges from $5\%$ to $20\%$ at medium and high values of the BDT score, respectively.
Alternatively, using two MC samples with different parton shower models,  $\powhegpythiaE$ and $\powhegherwig$, it was verified that for \wzjjqcd events the variations of the BDT score shape due to different parton shower models are within the global modelling uncertainty defined above. 
A global modelling uncertainty in the \wzjjew signal template is also estimated by comparing predictions of the BDT score distribution in the signal region from the \sherpa and \madgraph MC event generators.
This modelling uncertainty affects the shape of the BDT score distribution by at most $14\%$ at large values of the BDT score.
The Sherpa \wzjjew sample used in this analysis was recently found to implement a colour flow computation in VBS-like processes that increases central parton emissions from the parton shower~\cite{ATL-PHYS-PUB-2019-004}.
It was verified that possible effects on kinematic distributions and especially on the BDT score distribution are covered by the modelling uncertainty used.
%
%
The interference between electroweak- and QCD-induced processes is not included in the probability distribution functions of the fit but is considered as an uncertainty affecting only the shape of the \wzjjew MC template.
The effect is determined using the \madgraph MC generator, resulting for the signal region in shape-only uncertainties ranging from $10\%$ to $5\%$ at low and high values of the BDT score, respectively.
The effect of interference on the shape of the \wzjjew MC template in the \wzjj-QCD CR is negligible and is therefore not included.

Systematic uncertainties affecting the reconstruction and energy calibration of jets, electrons and muons are propagated through the analysis.
The dominant sources of uncertainties are the jet energy scale calibration, including the modelling of pile-up.
The uncertainties in the jet energy scale are obtained from $\sqrt{s} = 13$~\TeV\ simulations and {\textit{in situ}} measurements~\cite{Aaboud:2017jcu}. 
The uncertainty in the jet energy resolution~\cite{PERF-2011-04} and in the suppression of jets originating from pile-up are also considered~\cite{Aad:2015ina}.
The uncertainties in the $b$-tagging efficiency and the mistag rate are also taken into account.
The effect of jet uncertainties on the expected number of events ranges from $10\%$ to $3\%$ at low and high values of the BDT score, respectively, with a similar effect for \wzjjqcd and \wzjjew events.

 The uncertainty in the \MET measurement is estimated by propagating the uncertainties in the transverse momenta of hard physics objects and
 by applying momentum scale and resolution uncertainties to the track-based soft term~\cite{Aaboud:2018tkc,ATLAS-CONF-2018-023}. 

The uncertainties due to lepton reconstruction, identification, isolation requirements and trigger efficiencies are estimated using tag-and-probe methods in $Z \to \ell \ell$ events~\cite{ATLAS-CONF-2016-024,PERF-2015-10}.
Uncertainties in the lepton momentum scale and resolution are also assessed using  $Z \to \ell \ell$ events~\cite{PERF-2015-10,Aaboud:2018ugz}.
These uncertainties impact the expected number of  events by $1.4\%$ and $0.4\%$ for electrons and muons, respectively, and are independent of the BDT score.
Their effect is similar for \wzjjqcd and \wzjjew events.

A $40\%$ yield uncertainty is assigned to the reducible background estimate.
This takes into account the limited number of events in the control regions as well as the differences in background composition between the control regions used to determine the lepton misidentification rate and the control regions used to estimate the yield in the signal region.
The uncertainty due to irreducible background sources other than \wzjjqcd is evaluated by propagating the uncertainty in their MC cross-sections. These are $20\%$ for $VVV$~\cite{ATL-PHYS-PUB-2016-002}, $15\%$ for $tZj$~\cite{ATLASWZ_8TeV} and $t\bar{t}+V$~\cite{ATL-PHYS-PUB-2016-005}, and $25\%$ for \zzjjqcd to account for the potentially large impact of scale variations.

The uncertainty in the combined $2015$+$2016$ integrated luminosity is $2.1\%$. It is derived, following a methodology similar to that detailed in Ref.~\cite{DAPR-2013-01}, and using the LUCID-2 detector for the baseline luminosity measurements~\cite{LUCID2}, from a calibration of the luminosity scale using $x$--$y$ beam-separation scans.

The effect of the systematic uncertainties on the final results after the maximum-likelihood fit is shown in Table~\ref{table:sysOnMu_summary} where the breakdown of the
contributions to the uncertainties in the measured fiducial cross-section $\sigma^{\mathrm{fid.}}_{\wzjjew}$ is presented.
The individual sources of systematic uncertainty  are combined into theory modelling and experimental categories.
As shown in the table, the systematic uncertainties in the jet reconstruction and calibration play a dominant role, followed by the uncertainties in the modelling of the \wzjjew signal and of the \wzjjqcd background. 
Systematic uncertainties in the missing transverse momentum computation arise directly from the momentum and energy calibration of jets, electrons and muons and are included in the respective lines of Table~\ref{table:sysOnMu_summary}.   
Systematic uncertainties in the modelling of the reducible and irreducible backgrounds other than \wzjjqcd are also detailed.

\begin{table}[!htbp]
\caption{Summary of the relative uncertainties in the measured fiducial cross-section $\sigma^{\mathrm{fid.}}_{\wzjjew}$. The uncertainties are reported as percentages.
\label{table:sysOnMu_summary}}
\centering
\begin{tabular}{lc}
\toprule
Source & Uncertainty [\%]\\
\midrule

%
\wzjjew theory modelling           &  4.8\\
\wzjjqcd theory modelling          &  5.2\\
\wzjjew and \wzjjqcd interference  &  1.9\\
\midrule
Jets                               &  6.6\\
Pile-up                             &  2.2\\
Electrons                          &  1.4\\
Muons                              &  0.4\\
$b$-tagging                        &  0.1\\
MC statistics                      &  1.9\\
Misid. lepton background           &  0.9\\
Other backgrounds                  &  0.8\\
\midrule
Luminosity                         &  2.1 \\
\midrule
Total Systematics                  & 10.9\\
\toprule
\end{tabular} 
\end{table}


\section{Cross-section measurements}
\label{sec:result}

The signal strength $\mu_{\wzjjew}$ and its uncertainty are determined with a profile-likelihood-ratio test statistic~\cite{frequentist}. 
Systematic uncertainties in the input templates are treated as nuisance parameters with an assumed Gaussian distribution.
The distributions of $m_{jj}$ in the $ZZ$-CR control region, of $N_{b-\textrm{jets}}$ in the $b$-CR, of $m_{jj}$ in the  \wzjjqcd control region and of the BDT score in the signal region, with background normalisations, signal normalisation and nuisance parameters adjusted by the profile-likelihood fit are shown in Figure~\ref{fig:BDTScore}. 
The corresponding post-fit yields are detailed in Table~\ref{tab:VBSyield_postfit}.
The table presents the integral of the BDT score distribution in the SR, but the uncertainty on the measured signal cross section is dominated by events at high BDT score.
The signal strength is measured to be

\[
        \mu_{\wzjjew} = 1.77 \;^{+ 0.44} _{- 0.40}  \, \mathrm{(stat.)}  \;^{+ 0.15} _{- 0.12} \, \mathrm{(exp.\, syst.)} \;^{+ 0.15} _{- 0.12} \, \mathrm{(mod.\, syst.)} \;^{+ 0.15} _{- 0.13} \, \mathrm{(theory)} \;^{+ 0.04} _{- 0.02} \, \mathrm{(lumi.)}= 1.77 \;^{+ 0.51} _{- 0.45} \; , 
\label{eq:muEW_result}
\]


where the uncertainties correspond to statistical, experimental systematic, theory modelling and interference systematic, theory $\sigma^{\mathrm{fid., \, MC}}_{\wzjjew}$ normalisation and luminosity uncertainties, respectively.
The background-only hypothesis is excluded with a significance of $5.3$ standard deviations, compared with $3.2$  standard deviations expected.
The normalisation parameters of the \wzjjqcd, $t\bar{t}+V $ and $ZZ$ backgrounds constrained by data in the control and signal regions are measured to be  $\mu_{\wzjjqcd} = 0.56 \pm 0.16$, $\mu_{t\bar{t}+V} = 1.07 \pm 0.28$ and $\mu_{\zzjjqcd} = 1.34 \pm 0.44$. 
The observed \wzjjew production integrated fiducial cross-section derived from this signal strength for a single leptonic decay mode is

\[
\begin{array}{rcl}
\sigma^{\mathrm{fid.}}_{\wzjjew} &=& 0.57 \; ^{+ 0.14} _{- 0.13} \,\mathrm{(stat.)} \; ^{+ 0.05} _{- 0.04} \,\mathrm{(exp.\, syst.)} \; ^{+ 0.05} _{- 0.04} \,\mathrm{(mod.\, syst.)}  \; ^{+ 0.01} _{- 0.01} \,\mathrm{(lumi.)} \; \mathrm{fb} \, \\
        &=& 0.57 \; ^{+ 0.16} _{- 0.14} \; \mathrm{fb} \; .\\
\end{array}
 \]

%
It corresponds to the cross-section of electroweak \wzjj production, including interference effects between \wzjjqcd and \wzjjew processes, in the fiducial phase space defined in Section~\ref{sec:phase_space} using dressed-level leptons.

The SM LO prediction from \sherpa for electroweak production without interference effects is 
\[
\sigma^{\mathrm{fid., \, Sherpa}}_{\wzjjew} = 0.321 \pm 0.002 \,\mathrm{(stat.)} \pm 0.005 \,\mathrm{(PDF)} ^{+0.027}_{-0.023} \,\mathrm{(scale)}  \; \mathrm{fb},
\]
where the effects of uncertainties in the PDF and the $\alphas$ value used in the PDF determination, as well as the uncertainties due to the renormalisation and factorisation scales, are evaluated using the same procedure as the one described in Section~\ref{sec:systematics}.

A larger cross-section of $\sigma^{\mathrm{fid., \, MadGraph}}_{\wzjjew} = 0.366 \pm 0.004 \,\mathrm{(stat.)}  \; \mathrm{fb}$ is predicted by \madgraph.
These predictions are at \LO only and include neither the effects of interference, estimated at \LO to be $10\%$, nor the effects of NLO electroweak corrections as calculated recently in Ref.~\cite{Denner:2019tmn}.

\begin{figure}[!htbp]
\begin{center}
\includegraphics[width=0.4\textwidth]{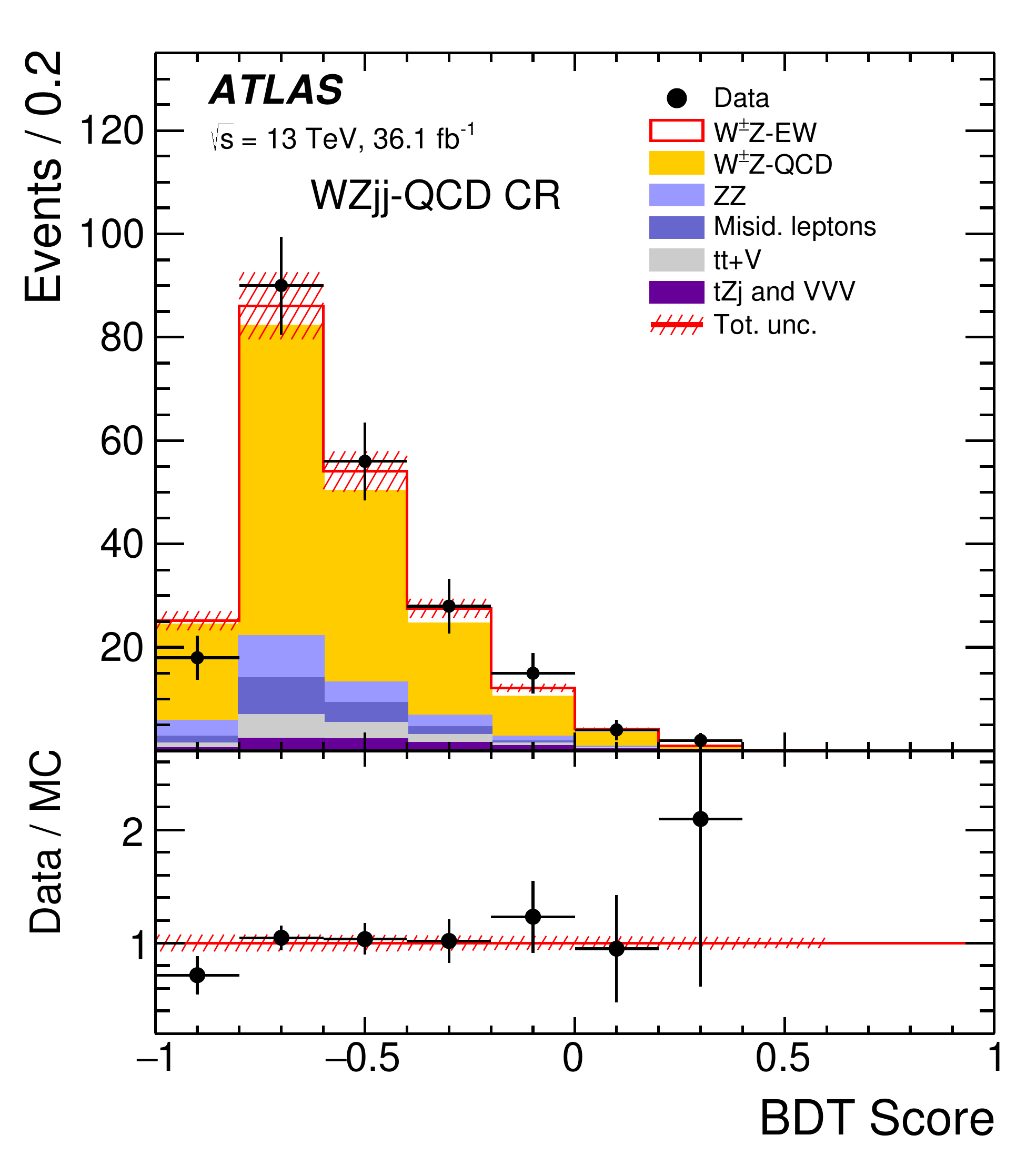}
\caption{Post-fit distribution of the BDT score distribution in the  \wzjjqcd control region. 
Signal and backgrounds are normalised to the expected number of events after the fit. 
The uncertainty band around the MC expectation includes all systematic uncertainties as obtained from the fit.
}
\label{fig:BDTScore_WZjjQCDCR}
\end{center}
\end{figure}

\begin{figure}[!htbp]
\begin{center}
\includegraphics[width=0.4\textwidth]{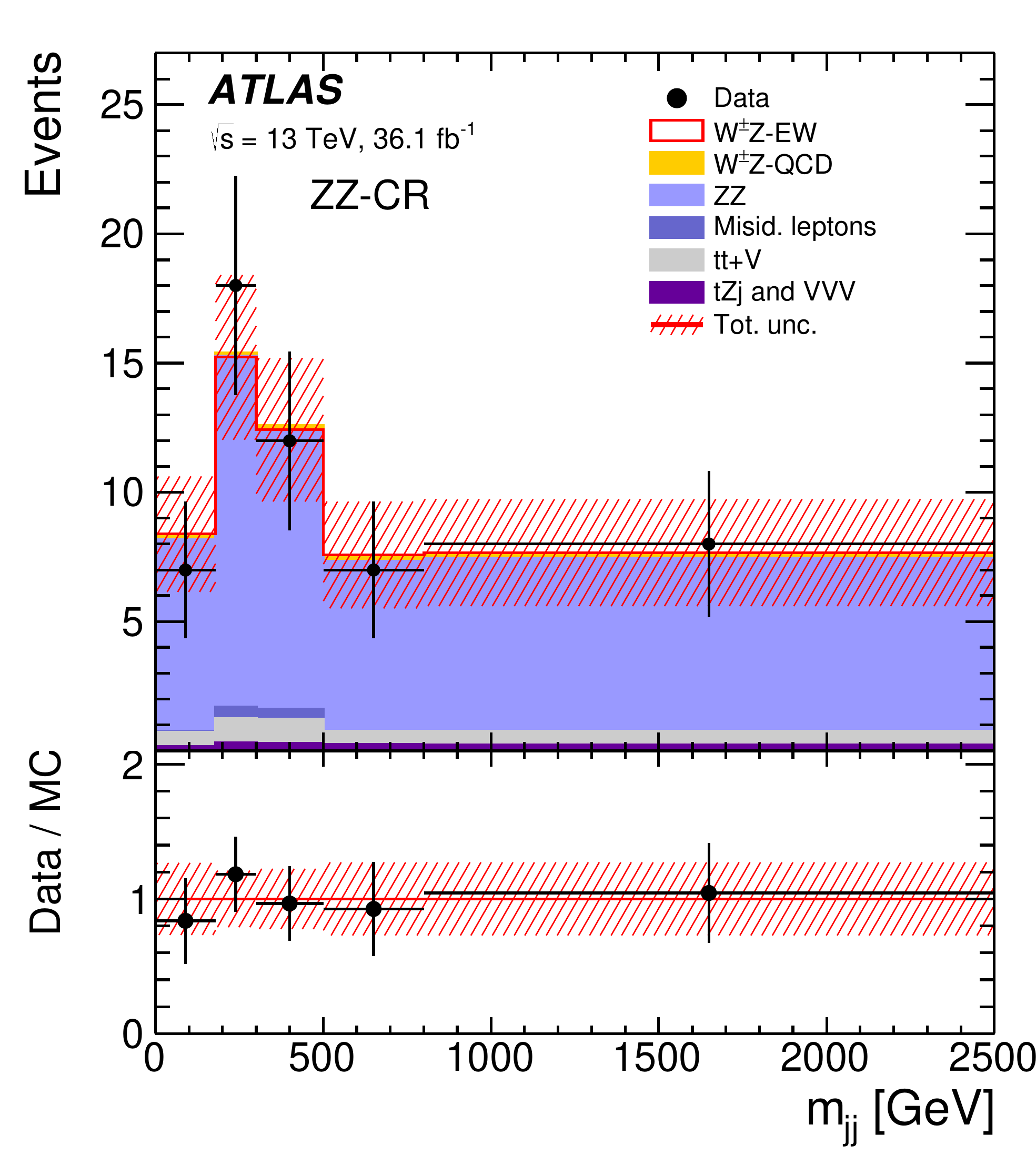}\put(-40,130){{(a)}}
\includegraphics[width=0.4\textwidth]{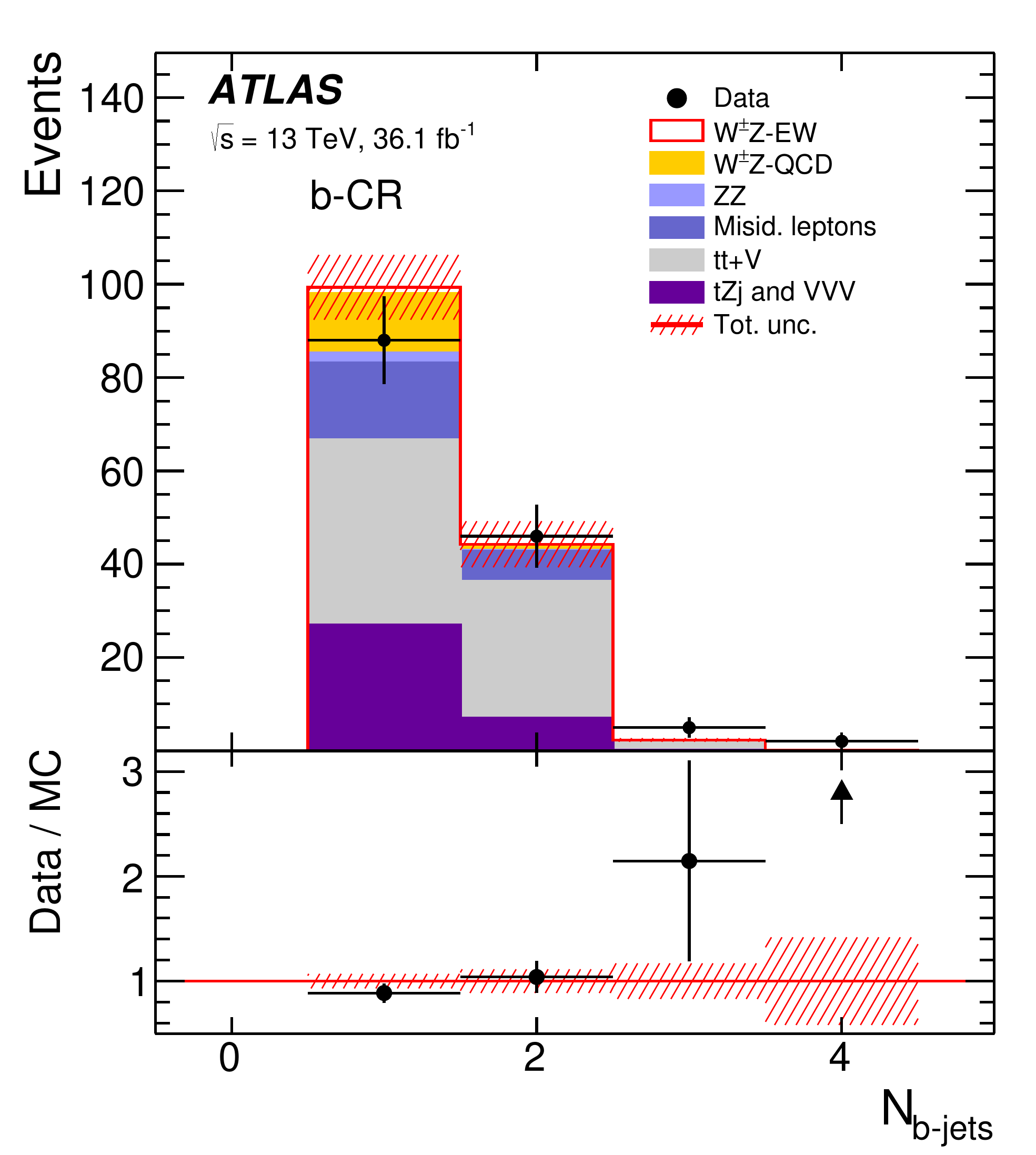}\put(-40,130){{(b)}}\\
\includegraphics[width=0.4\textwidth]{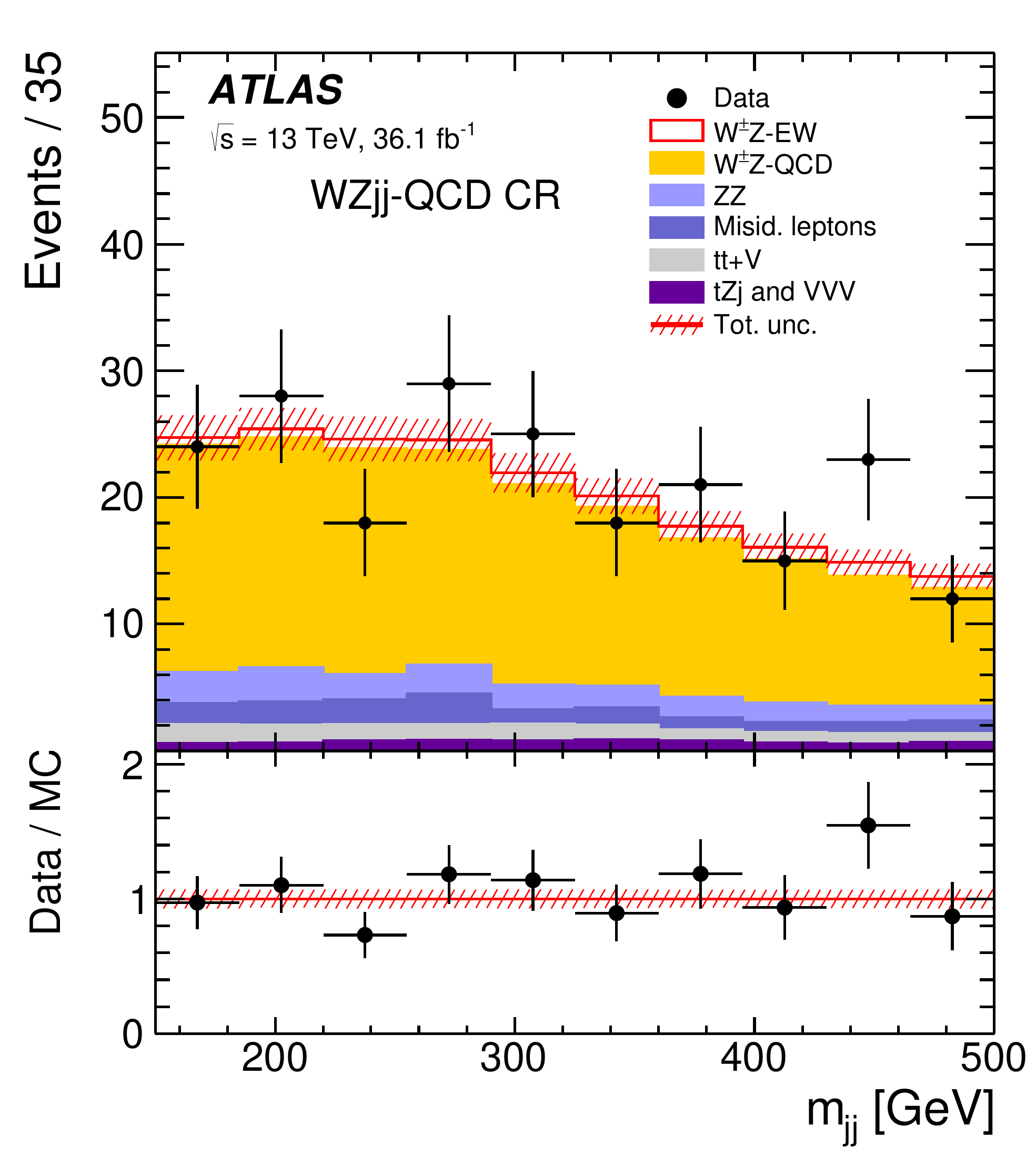}\put(-50,130){{(c)}}
\includegraphics[width=0.4\textwidth]{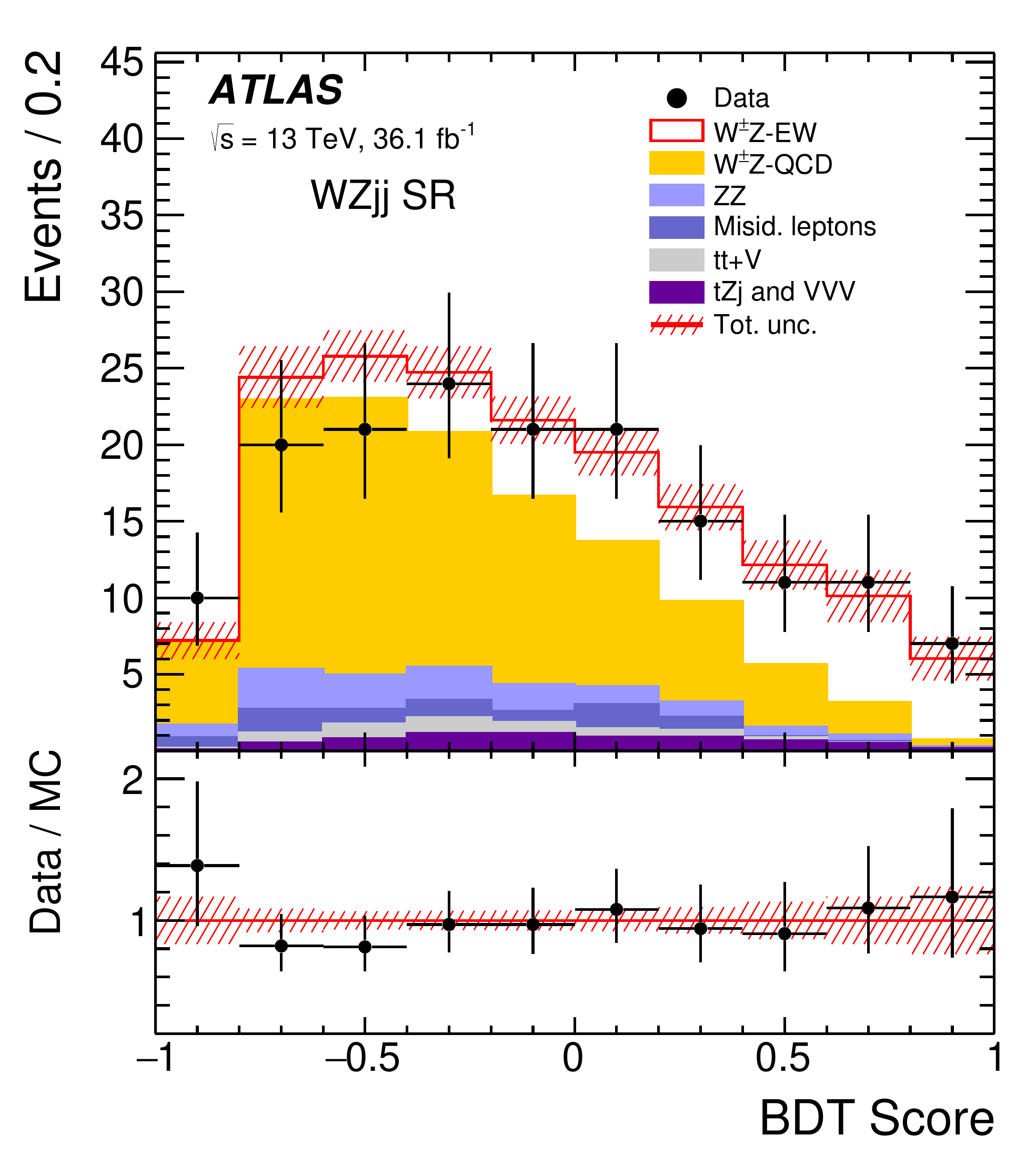}\put(-40,130){{(d)}}
\caption{Post-fit distributions of (a) $m_{jj}$ in the $ZZ$-CR control region, (b) $N_{b-\textrm{jets}}$ in the $b$-CR, (c) $m_{jj}$ in the  \wzjjqcd control region and (d) the BDT score distribution in the signal region. 
Signal and backgrounds are normalised to the expected number of events after the fit. 
The uncertainty band around the MC expectation includes all systematic uncertainties as obtained from the fit.
}
\label{fig:BDTScore}
\end{center}
\end{figure}

\begin{table}[!htbp] 
\caption{Observed and expected numbers of events in the \wzjj signal region and in the three control regions, after the fit.
The expected number of \wzjjew events from \sherpa and the estimated number of background events from the other processes are shown.
The sum of the backgrounds containing misidentified leptons is labelled `Misid. leptons'.
The total correlated post-fit uncertainties are quoted.
\label{tab:VBSyield_postfit}}
\centering
\begin{tabular}{%
l
S@{$\pm$}S
S@{$\pm$}S
S@{$\pm$}S
S@{$\pm$}S
} 
\toprule
                      &  \multicolumn{2}{c}{SR}		   &  \multicolumn{2}{c}{\wzjjqcd CR}	        &  \multicolumn{2}{c}{$b$-CR}	    & \multicolumn{2}{c}{$ZZ$-CR}           \\
\midrule 
Data                  & \multicolumn{2}{c}{$ 161$}             & \multicolumn{2}{c}{$ 213$}             & \multicolumn{2}{c}{$ 141$}            & \multicolumn{2}{c}{$52$}              \\
Total predicted       &  167   & 11    &  204   & 12    &  146  & 11    & 51.3  & 7.0    \\  
\midrule                                                                          
                                                                                
\wzjjew  (signal)     &  44    & 11    &  8.52  & 0.41  &  1.38 & 0.10  &  0.211& 0.004 \\                          
\wzjjqcd              &  91    & 10    &  144   & 14    &  13.9 & 3.8   &  0.94 & 0.14  \\
Misid. leptons        &  7.8   & 3.2   &  14.0  & 5.7   &  23.5 & 9.6   &  0.41 & 0.18  \\
\zzjjqcd	      &  11.1  & 2.8   &  18.3  & 1.1   &  2.35 & 0.06  &  40.8 & 7.2   \\
$tZj$		      &  6.2   & 1.1   &  6.3   & 1.1   &  34.0 & 5.3   &  0.17 & 0.04  \\
$t\bar{t}+V$	      &  4.7   & 1.0   &  11.14 & 0.37  &  71   & 15    &  3.47 & 0.54  \\
\zzjjew 	      &  1.80  & 0.45  &  0.44  & 0.10  &  0.10 & 0.03  &  4.2  & 1.2   \\
$VVV$  	     	      &  0.59  & 0.15  &  0.93  & 0.23  &  0.13 & 0.03  &  1.06 & 0.30  \\
\bottomrule

\end{tabular}
\end{table}

From the number of observed events in the SR, the integrated cross-section of \wzjj production in the VBS fiducial phase space defined in Section~\ref{sec:phase_space}, including \wzjjew and \wzjjqcd contributions and their interference, is measured.
For a given channel \wz\ $\rightarrow \ell^{'\pm} \nu \ell^+ \ell^-$, where $\ell$ and $\ell^{'}$ indicates each type of lepton ($e$ or $\mu$),  the integrated fiducial cross section that includes the leptonic branching fractions of the $W$ and $Z$ bosons is calculated as
 \[
\sigma^{\mathrm{fid.}}_{\wzjj}  = \frac {  N_{\textrm{data}} - N_{\textrm{bkg}} }
 {   \mathcal{L}  \cdot  C_{WZjj} }  \times  \left( 1 - \frac {N_\tau} {N_{\textrm{all}} } \right) \, ,
 \]
 where  $N_{\textrm{data}}$ and  $N_{\textrm{bkg}}$ are the number of observed events and the estimated number of background events in the SR, respectively, 
and $\mathcal{L}$ is the integrated luminosity. 
The factor  $C_{WZjj}$, obtained from simulation, is the ratio of the number of selected signal events at detector level to the number of events at particle level in the fiducial phase space. 
This factor corrects for detector efficiencies and for QED final-state radiation effects. 
The contribution from $\tau$-lepton decays, amounting to $4.7\%$, is removed from the cross-section definition by introducing the term in parentheses.
This term is computed using simulation, where $N_\tau$ is the number of selected events at detector level in which at least one of the bosons decays into a $\tau$-lepton and $N_{\textrm{all}}$ is the number of selected  $WZ$ events with decays into any lepton.
The $C_{WZjj}$ factor calculated with \sherpa for the sum of the four measured decay channels is $0.52$ with a negligible statistical uncertainty.
This factor is the same for \wzjjqcd and \wzjjew events, as predicted by \sherpa.
The theory modelling uncertainty in this factor is $8\%$, as estimated from the difference between the \sherpa and \madgraph predictions.
The uncertainties on this factor due to higher order QCD scale corrections or PDF are negligible.

The measured \wzjj cross-section in the fiducial phase space for a single leptonic decay mode  is

\begin{eqnarray*}
\sigma_{W^{\pm}Zjj}^{\mathrm{fid.}} & = & 1.68  \pm 0.16 \,\mathrm{(stat.)} \pm 0.12 \,\mathrm{(exp.\, syst.)}  \pm 0.13 \,\mathrm{(mod.\, syst.)} \pm 0.044 \,\mathrm{(lumi.)} \; \mathrm{fb} \, , \nonumber\\
                                    & = & 1.68  \pm 0.25 \; \mathrm{fb} \, ,
\end{eqnarray*}

where the uncertainties correspond to statistical, experimental systematic, theory modelling systematic, and luminosity uncertainties, respectively.
The corresponding prediction from \sherpa for strong and electroweak production without interference effects is 
\[
\sigma^{\mathrm{fid.,\, Sherpa}}_{\wzjj} = 2.15 \pm 0.01 \,\mathrm{(stat.)} \pm 0.05 \,\mathrm{(PDF)} ^{+0.65}_{-0.44} \,\mathrm{(scale)}  \; \mathrm{fb}.
\]


Events in the SR are also used to  measure the \wzjj\ differential production cross-section in the VBS fiducial phase space. 
The differential detector-level distributions are corrected for detector resolution %
 using an iterative Bayesian unfolding method~\cite{DAgostini:1994zf}, as implemented in the RooUnfold toolkit~\cite{RooUnfold}.
Three iterations were used for the unfolding of each variable.
The width of the bins in each distribution is chosen according to the experimental resolution and to the statistical significance of the expected number of events in that bin.
The fraction of signal MC events reconstructed in the same bin as generated is always greater than $40\%$ and around $70\%$ on average.

For each  distribution, simulated \wzjj\ events are used to obtain a response matrix that accounts for bin-to-bin migration effects between the reconstruction-level and particle-level distributions.
The \sherpa MC samples for \wzjjew\ and \wzjjqcd\ production are added together to model \wzjj production. 
To more closely model the data and to minimise unfolding uncertainties, their predicted cross-sections are rescaled by the respective signal strengths of $1.77$ and $0.56$ for the \wzjjew\ and \wzjjqcd\ contributions, respectively, as measured in data by the maximum-likelihood fit.

Uncertainties in the unfolding due to imperfect modelling of the data by the
MC simulation are evaluated using a data-driven method~\cite{Malaescu:2011yg}, where the MC differential distribution is corrected to
match the data distribution and the resulting weighted MC distribution at reconstruction level is unfolded
with the response matrix used in the data unfolding. 
The new unfolded distribution is compared with the weighted MC distribution at generator level and the difference is taken as the systematic uncertainty.
The uncertainties obtained range from $0.1\%$ to $25\%$ depending on the resolution of the unfolded observables and on the quality of its description by \sherpa.

Measurements are performed as a function of three variables sensitive to anomalies in the quartic gauge coupling in \wzjj events~\cite{ATLASWZ_8TeV}, namely the scalar sum of the transverse momenta of the three charged leptons associated with the $W$ and $Z$ bosons \sptl, the difference in azimuthal angle \dPhiWZ\ between the $W$ and $Z$ bosons' directions, and the transverse mass of the \wz system \mtwz, defined following Ref.~\cite{ATLASWZ_8TeV}.
These are presented in Figure~\ref{fig:diffXSection:SptL_dPhiWZ}.
Measurements are also performed as a function of variables related to the kinematics of jets. 
The exclusive multiplicity of jets, \njets, is shown  in Figure~\ref{fig:diffXSection:Njets}.
The absolute difference in rapidity between the two tagging jets \dYjj, the invariant mass of the tagging jets \mjj, the exclusive multiplicity \NjGap of jets with \mbox{$p_\textrm{T} > 25$~\GeV}\ in the gap in $\eta$ between the two tagging jets, and the azimuthal angle between the two tagging jets \dPhijj are shown in Figure~\ref{fig:diffXSection:dYjj_mjj}.

Total uncertainties in the measurements are dominated by statistical uncertainties.
The differential measurements are compared with the prediction from \sherpa, after having rescaled the separate \wzjjqcd and \wzjjew components by the global $\mu_{\wzjjqcd}$ and $\mu_{\wzjjew}$ parameters, respectively, obtained from the profile-likelihood fit to data.
Interference effects between the \wzjjqcd and \wzjjew processes are incorporated via the $\mu_{\wzjjew}$ parameter as a change of the global normalisation of the \sherpa electroweak prediction.

\begin{figure}[!htbp]
\begin{center}
\includegraphics[width=0.4\textwidth]{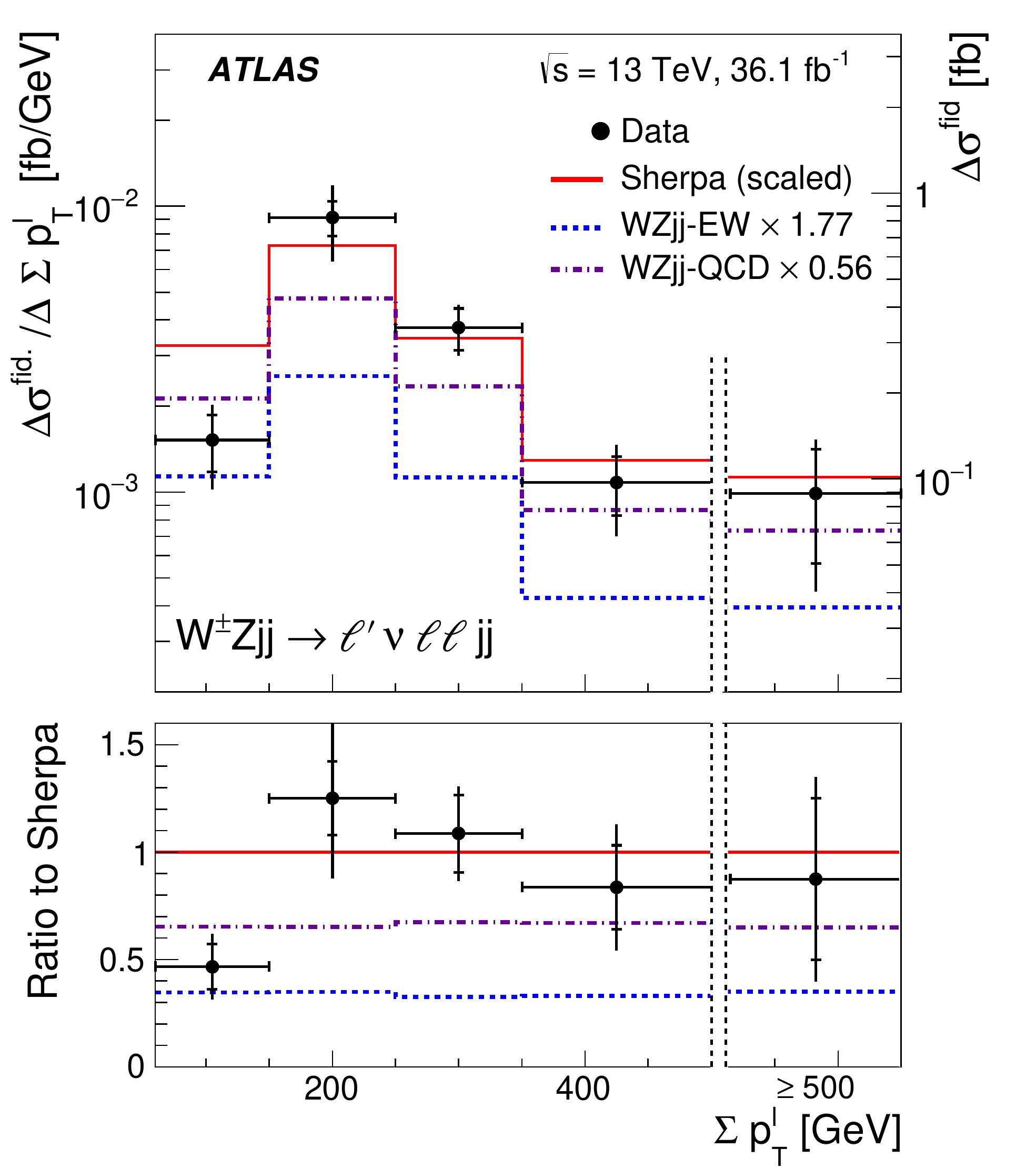}\put(-150,180){{(a)}}
\includegraphics[width=0.4\textwidth]{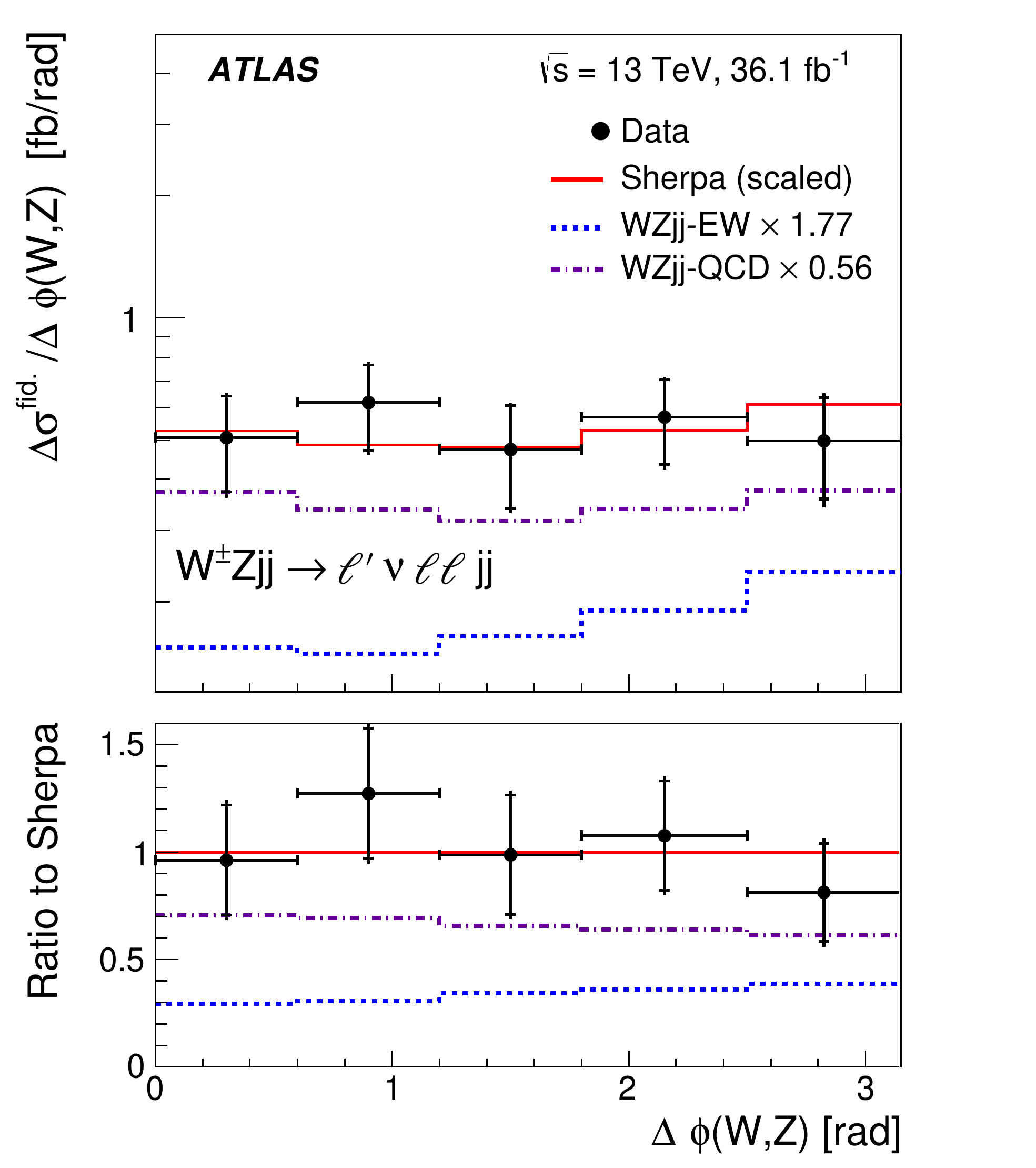}\put(-150,180){{(b)}}\\
\includegraphics[width=0.4\textwidth]{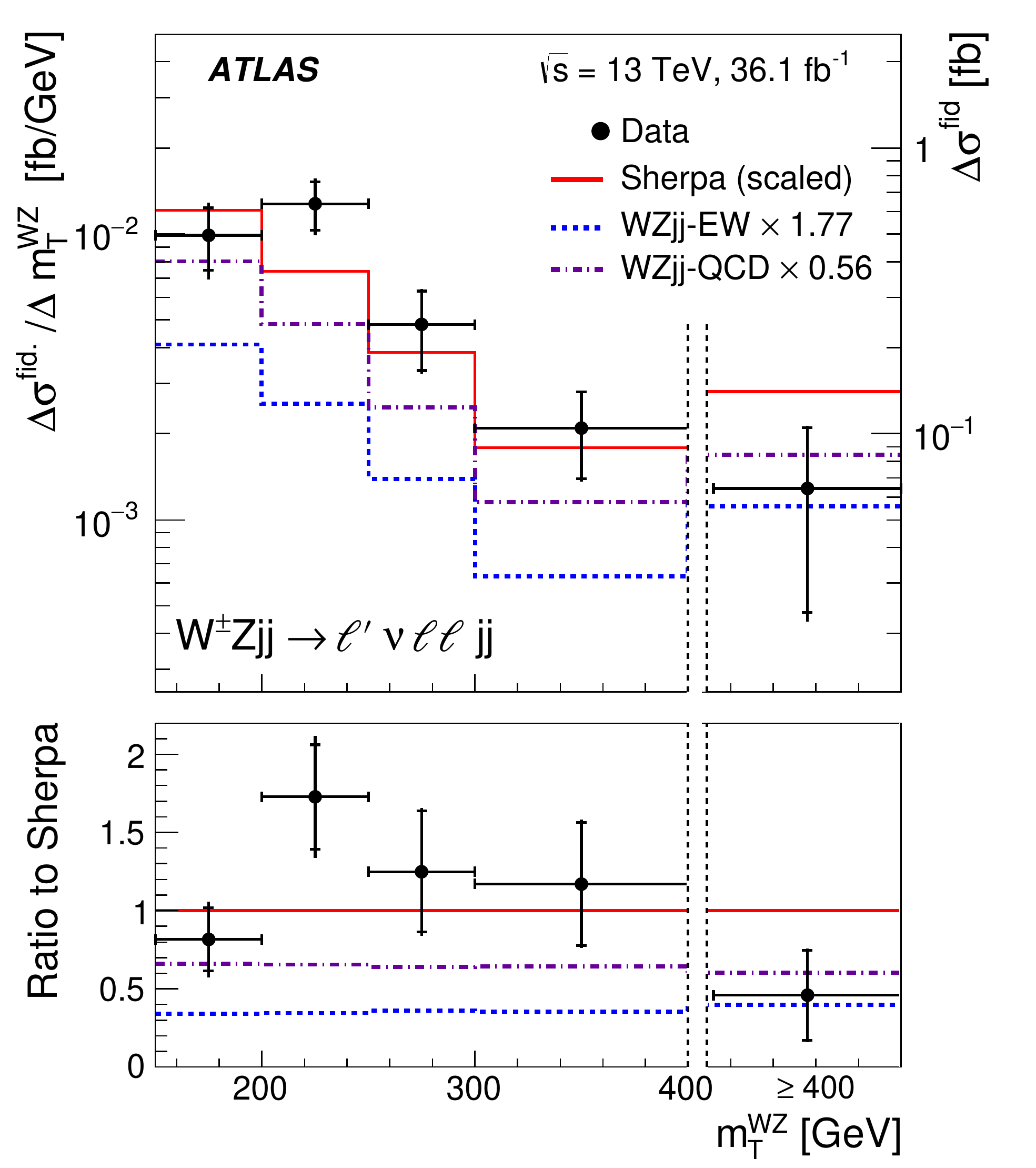}\put(-150,180){{(c)}}

\caption{The measured \wzjj differential cross-section in the VBS fiducial phase space as a function of (a) \sptl, (b) \dPhiWZ and (c) \mtwz.
The inner and outer error bars on the data points represent the statistical and total uncertainties, respectively.
The measurements are compared with the sum of the rescaled \wzjjqcd and \wzjjew predictions from \sherpa (solid line).
The \wzjjew and \wzjjqcd contributions are also represented by dashed and dashed-dotted lines, respectively.
In (a) and (c), the right $y$-axis refers to the last cross-section point, separated from the others by a vertical dashed line, as this last bin is integrated up to the maximum value reached in the phase space. 
The lower panels show the ratios of the data to the predictions from \sherpa. 
The uncertainty on the \sherpa prediction is dominated by the QCD scale uncertainty on the \wzjjqcd predicted cross-section, whose envelope is of $^{+30}_{-20}\%$ and it is not represented on the figure.
}
\label{fig:diffXSection:SptL_dPhiWZ}
\end{center}
\end{figure}

\begin{figure}[!htbp]
\begin{center}
\includegraphics[width=0.45\textwidth]{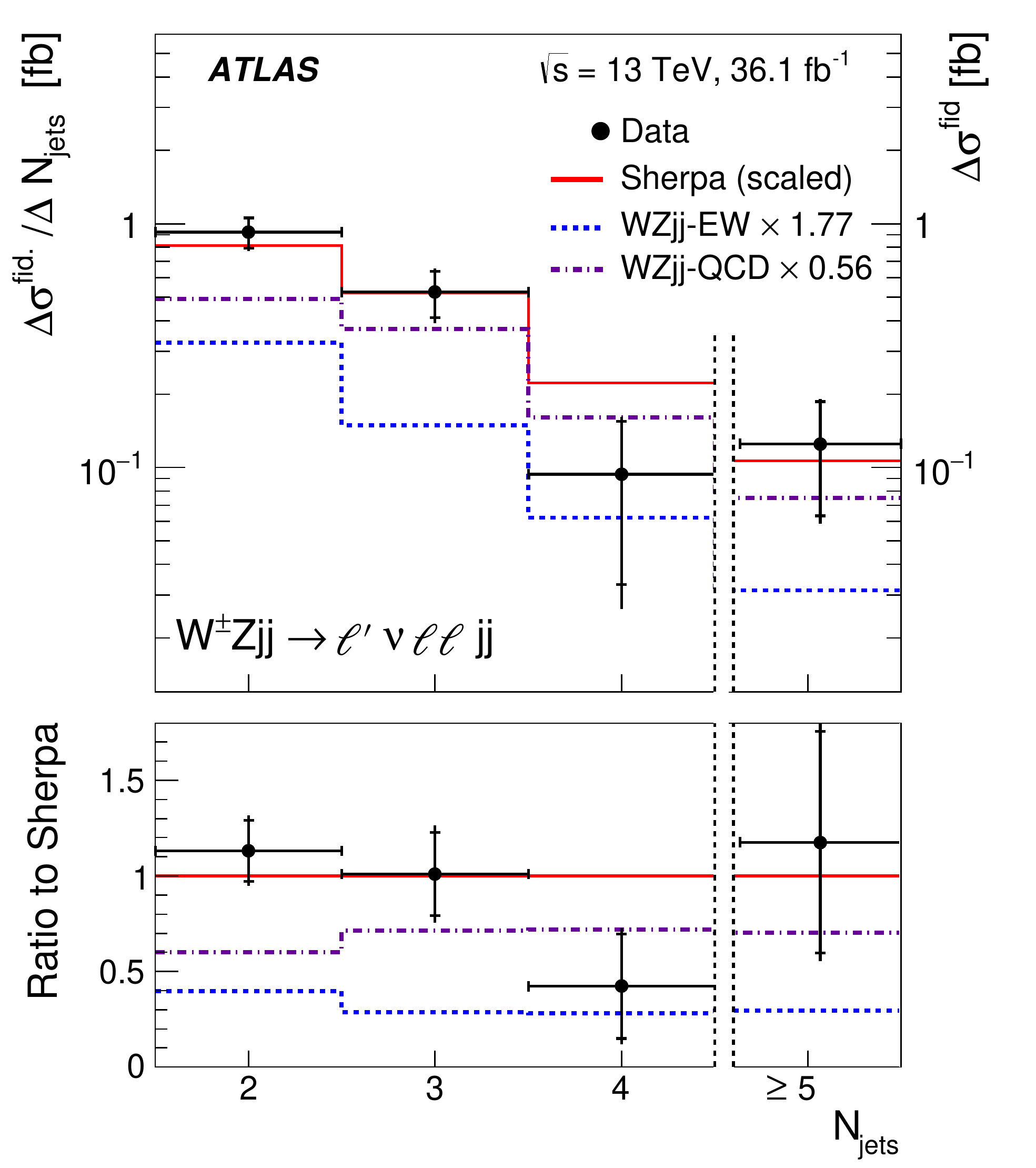}
\caption{The measured \wzjj differential cross-section in the VBS fiducial phase space as a function of  the exclusive jet multiplicity of jets with $p_\textrm{T} > 40$~\GeV.
The inner and outer error bars on the data points represent the statistical and total uncertainties, respectively.
The measurements are compared with the sum of the scaled \wzjjqcd and \wzjjew predictions from \sherpa (solid line).
The \wzjjew and \wzjjqcd contributions are also represented by dashed and dashed-dotted lines, respectively.
The right $y$-axis refers to the last cross-section point, separated from the others by a vertical dashed line, as this last bin is integrated up to the maximum value reached in the phase space.
The lower panel shows the ratio of the data to the prediction from \sherpa. 
The uncertainty on the \sherpa prediction is dominated by the QCD scale uncertainty on the \wzjjqcd predicted cross-section, whose envelope is of $^{+30}_{-20}\%$ and it is not represented on the figure.
}
\label{fig:diffXSection:Njets}
\end{center}
\end{figure}

\begin{figure}[!htbp]
\begin{center}
\includegraphics[width=0.4\textwidth]{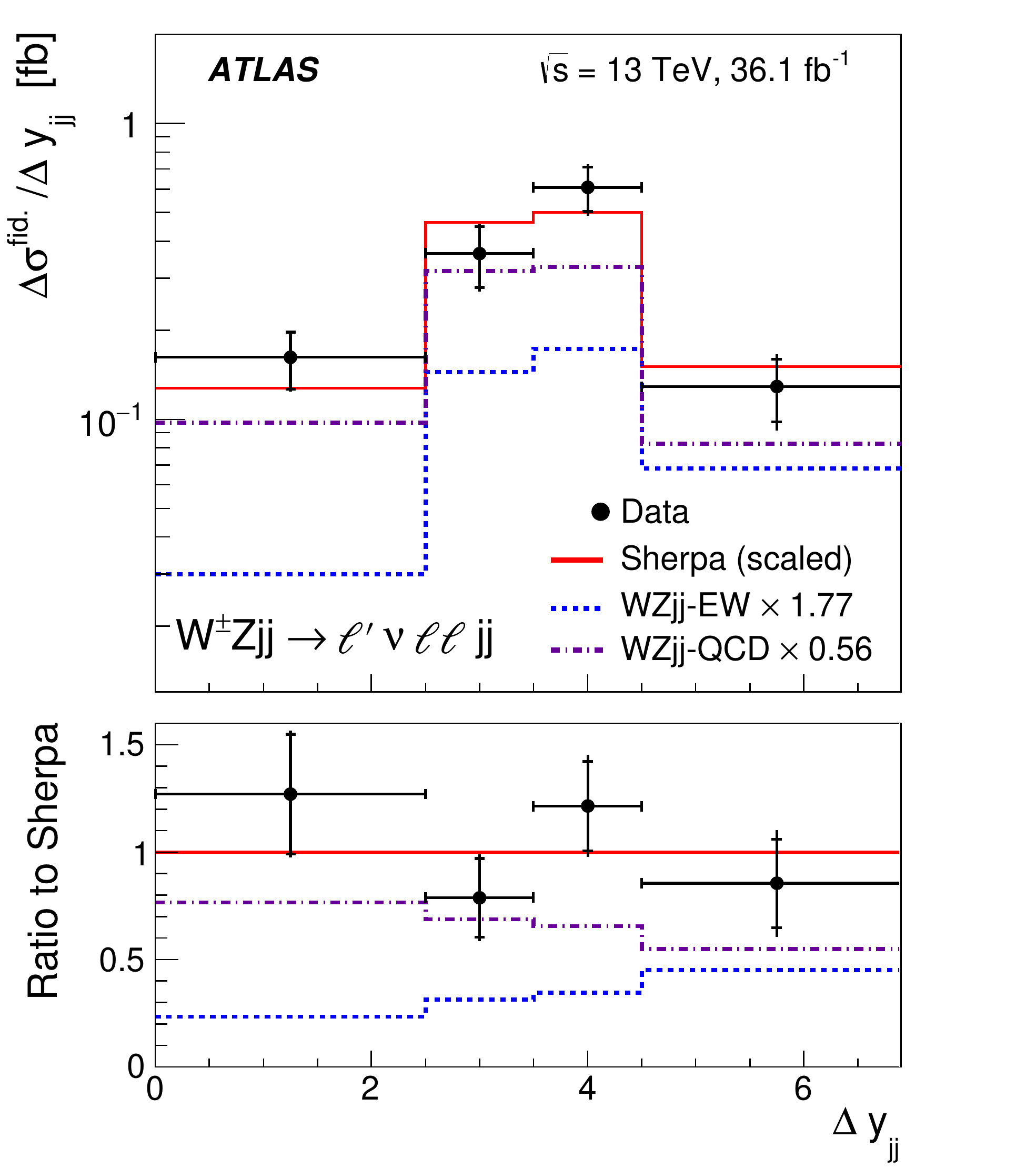}\put(-150,180){{(a)}}
\includegraphics[width=0.4\textwidth]{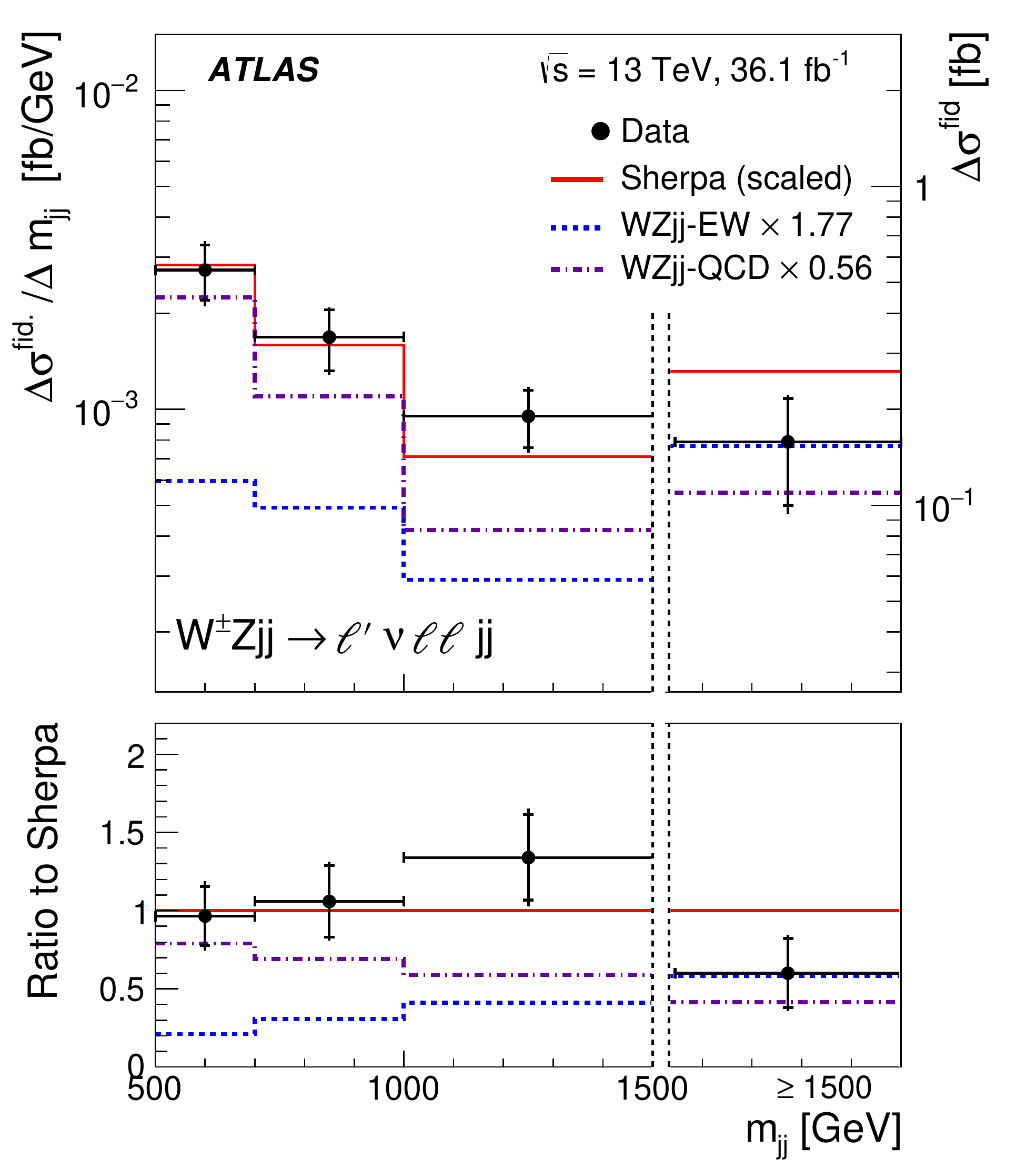}\put(-150,180){{(b)}}\\
\includegraphics[width=0.4\textwidth]{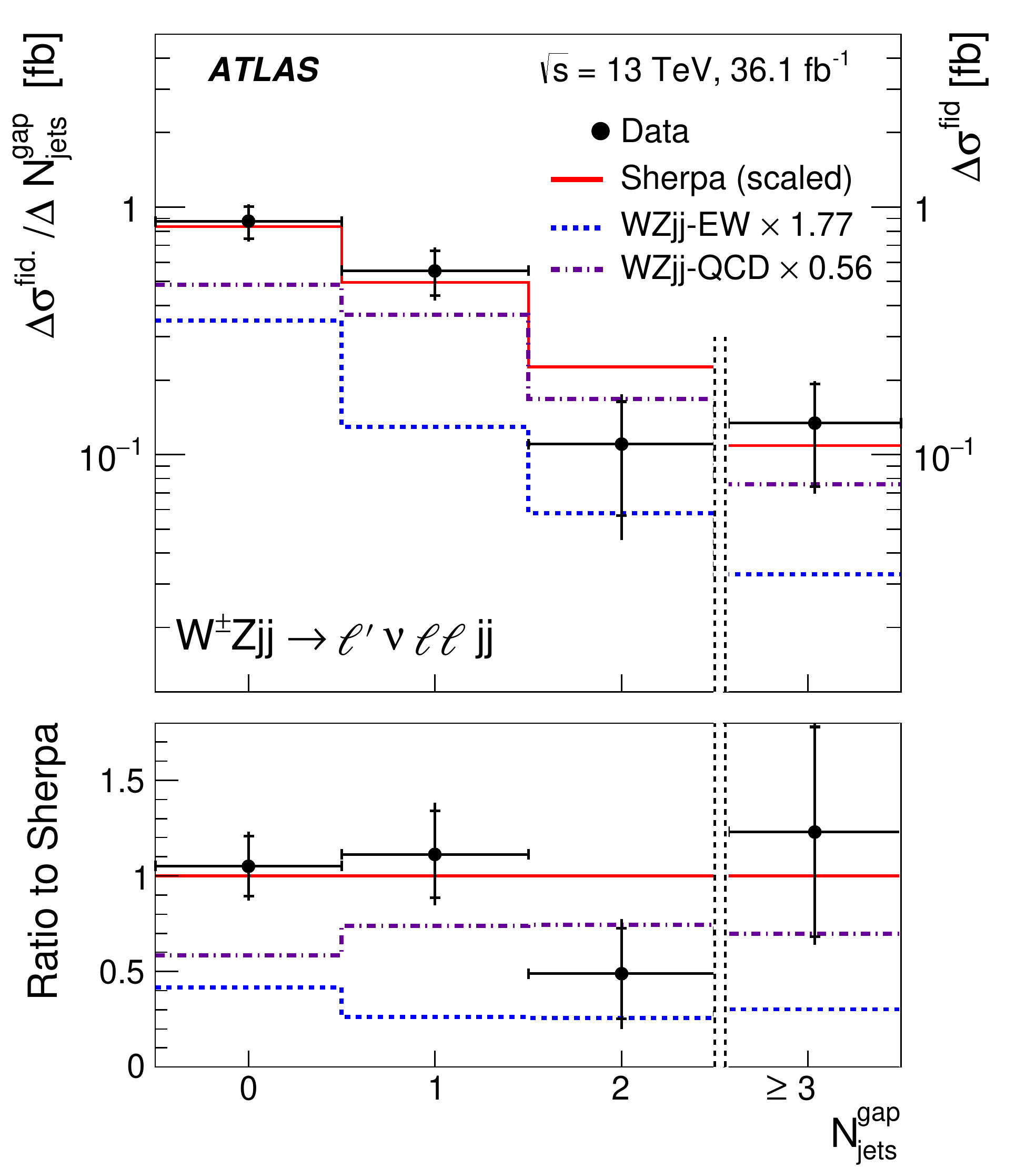}\put(-150,180){{(c)}}
\includegraphics[width=0.4\textwidth]{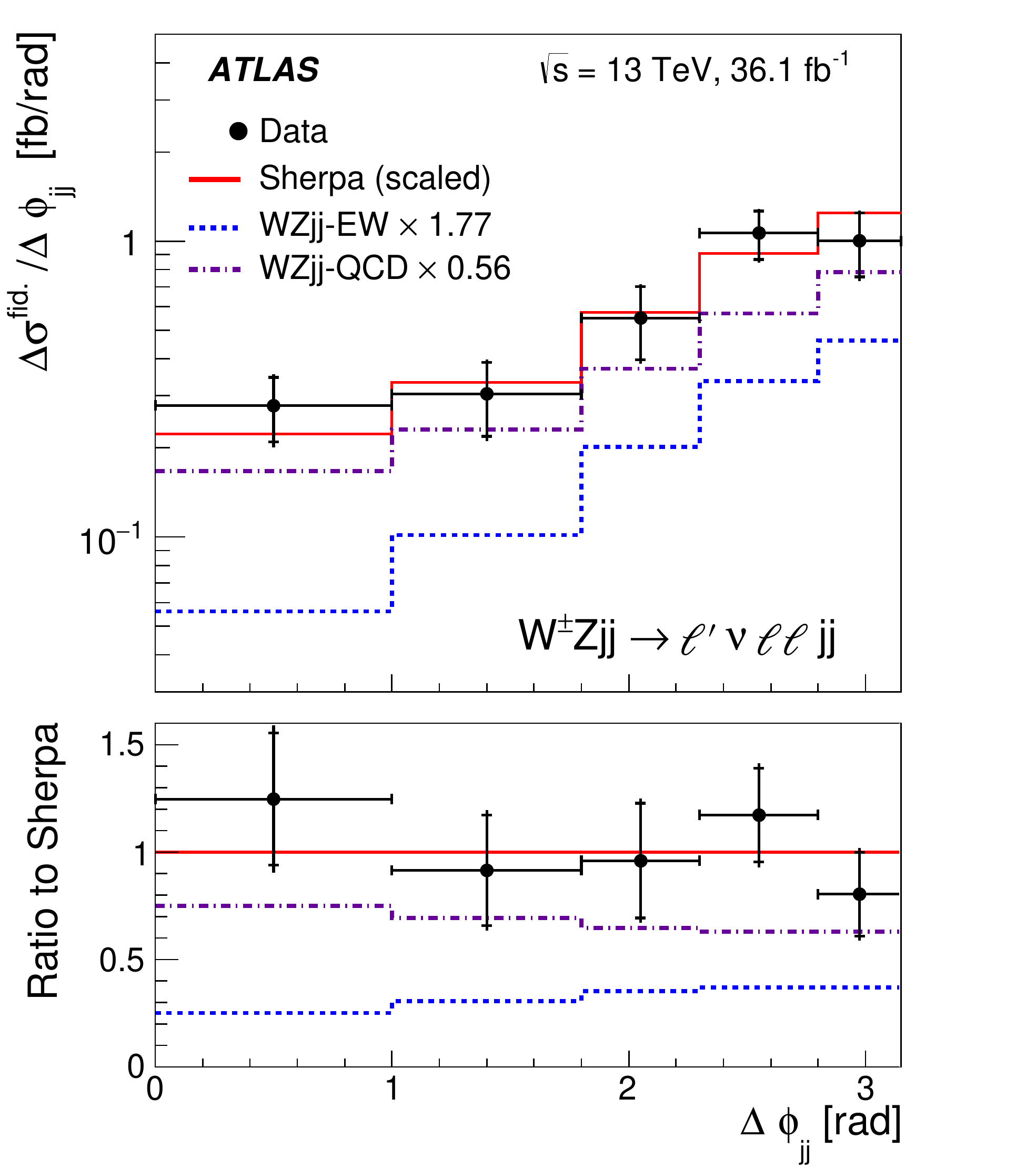}\put(-40,180){{(d)}}
\caption{The measured \wzjj differential cross-section in the VBS fiducial phase space as a function of (a) the absolute difference in rapidity between the two tagging jets \dYjj, (b) the invariant mass of the tagging jets \mjj, (c) \NjGap\, the exclusive jet multiplicity of jets with $p_\textrm{T} > 25$~\GeV\ in the gap between the two tagging jets, and (d) the azimuthal angle between the two tagging jets \dPhijj.
The inner and outer error bars on the data points represent the statistical and total uncertainties, respectively.
The measurements are compared with the sum of the rescaled \wzjjqcd and \wzjjew predictions from \sherpa (solid line).
The \wzjjew and \wzjjqcd contributions are also represented by dashed and dashed-dotted lines, respectively.
In (b) ad (c), the right $y$-axis refers to the last cross-section point, separated from the others by a vertical dashed line, as this last bin is integrated up to the maximum value reached in the phase space.
The lower panels show the ratios of the data to the predictions from \sherpa.
The uncertainty on the \sherpa prediction is dominated by the QCD scale uncertainty on the \wzjjqcd predicted cross-section, whose envelope is of $^{+30}_{-20}\%$ and it is not represented on the figure.
}
\label{fig:diffXSection:dYjj_mjj}
\end{center}
\end{figure}

\FloatBarrier

\section{Conclusion}
\label{sec:conclusion}

An observation of electroweak production of a diboson \wz\ system in association with two jets and measurements of its  production cross-section in $\sqrt{s} = 13$~\TeV\ $pp$ collisions at the LHC are presented.
The data were collected with the ATLAS detector and correspond to an integrated luminosity of $36.1~\ifb$.
The measurements use leptonic decays of the gauge bosons into electrons or muons and are performed in a fiducial phase space approximating the detector acceptance that increases the sensitivity to \wzjj\ electroweak production modes.
The electroweak production of \wz bosons in association with two jets is measured with observed and expected significances of $5.3$ and $3.2$ standard deviations, respectively.
The measured fiducial cross-section for electroweak production including interference effects is  
\[
\sigma_{\wzjjew} = 
0.57 \; ^{+ 0.14} _{- 0.13} \,(\mathrm{stat.}) \\ \; ^{+ 0.05} _{- 0.04} \,(\mathrm{exp.\, syst.}) \; ^{+ 0.05} _{- 0.04} \,(\mathrm{mod.\, syst.}) \; ^{+ 0.01} _{- 0.01} \,(\mathrm{lumi.})\; \mathrm{fb}.
\]
It is found to be larger than the \LO SM prediction of  $0.32 \pm 0.03$~fb as calculated with the \sherpa MC event generator that includes neither interference effects, estimated at \LO to be $10\%$, nor \NLO electroweak corrections.
Differential cross-sections of \wzjj production, including both the strong and electroweak processes, are also measured in the same fiducial phase space as a function of several kinematic observables.

\section*{Acknowledgements}


We thank CERN for the very successful operation of the LHC, as well as the
support staff from our institutions without whom ATLAS could not be
operated efficiently.

We acknowledge the support of ANPCyT, Argentina; YerPhI, Armenia; ARC, Australia; BMWFW and FWF, Austria; ANAS, Azerbaijan; SSTC, Belarus; CNPq and FAPESP, Brazil; NSERC, NRC and CFI, Canada; CERN; CONICYT, Chile; CAS, MOST and NSFC, China; COLCIENCIAS, Colombia; MSMT CR, MPO CR and VSC CR, Czech Republic; DNRF and DNSRC, Denmark; IN2P3-CNRS, CEA-DRF/IRFU, France; SRNSFG, Georgia; BMBF, HGF, and MPG, Germany; GSRT, Greece; RGC, Hong Kong SAR, China; ISF and Benoziyo Center, Israel; INFN, Italy; MEXT and JSPS, Japan; CNRST, Morocco; NWO, Netherlands; RCN, Norway; MNiSW and NCN, Poland; FCT, Portugal; MNE/IFA, Romania; MES of Russia and NRC KI, Russian Federation; JINR; MESTD, Serbia; MSSR, Slovakia; ARRS and MIZ\v{S}, Slovenia; DST/NRF, South Africa; MINECO, Spain; SRC and Wallenberg Foundation, Sweden; SERI, SNSF and Cantons of Bern and Geneva, Switzerland; MOST, Taiwan; TAEK, Turkey; STFC, United Kingdom; DOE and NSF, United States of America. In addition, individual groups and members have received support from BCKDF, CANARIE, CRC and Compute Canada, Canada; COST, ERC, ERDF, Horizon 2020, and Marie Sk{\l}odowska-Curie Actions, European Union; Investissements d' Avenir Labex and Idex, ANR, France; DFG and AvH Foundation, Germany; Herakleitos, Thales and Aristeia programmes co-financed by EU-ESF and the Greek NSRF, Greece; BSF-NSF and GIF, Israel; CERCA Programme Generalitat de Catalunya, Spain; The Royal Society and Leverhulme Trust, United Kingdom. 

The crucial computing support from all WLCG partners is acknowledged gratefully, in particular from CERN, the ATLAS Tier-1 facilities at TRIUMF (Canada), NDGF (Denmark, Norway, Sweden), CC-IN2P3 (France), KIT/GridKA (Germany), INFN-CNAF (Italy), NL-T1 (Netherlands), PIC (Spain), ASGC (Taiwan), RAL (UK) and BNL (USA), the Tier-2 facilities worldwide and large non-WLCG resource providers. Major contributors of computing resources are listed in Ref.~\cite{ATL-GEN-PUB-2016-002}.


%
%
%

\printbibliography

\clearpage 
 
\begin{flushleft}
{\Large The ATLAS Collaboration}

\bigskip

M.~Aaboud$^\textrm{\scriptsize 35d}$,    
G.~Aad$^\textrm{\scriptsize 100}$,    
B.~Abbott$^\textrm{\scriptsize 127}$,    
O.~Abdinov$^\textrm{\scriptsize 13,*}$,    
B.~Abeloos$^\textrm{\scriptsize 131}$,    
D.K.~Abhayasinghe$^\textrm{\scriptsize 92}$,    
S.H.~Abidi$^\textrm{\scriptsize 166}$,    
O.S.~AbouZeid$^\textrm{\scriptsize 40}$,    
N.L.~Abraham$^\textrm{\scriptsize 155}$,    
H.~Abramowicz$^\textrm{\scriptsize 160}$
% ...
\bigskip
\\

% ...
$^{13}$Institute of Physics, Azerbaijan Academy of Sciences, Baku; Azerbaijan.\\
% ...
$^{35}$$^{(a)}$Facult\'e des Sciences Ain Chock, R\'eseau Universitaire de Physique des Hautes Energies - Universit\'e Hassan II, Casablanca;$^{(b)}$Facult\'{e} des Sciences, Universit\'{e} Ibn-Tofail, K\'{e}nitra;$^{(c)}$Facult\'e des Sciences Semlalia, Universit\'e Cadi Ayyad, LPHEA-Marrakech;$^{(d)}$Facult\'e des Sciences, Universit\'e Mohamed Premier and LPTPM, Oujda;$^{(e)}$Facult\'e des sciences, Universit\'e Mohammed V, Rabat; Morocco.\\
% ...
$^{40}$Niels Bohr Institute, University of Copenhagen, Copenhagen; Denmark.\\
% ...
$^{92}$Department of Physics, Royal Holloway University of London, Egham; United Kingdom.\\
% ...
$^{100}$CPPM, Aix-Marseille Universit\'e, CNRS/IN2P3, Marseille; France.\\
% ...
$^{127}$Homer L. Dodge Department of Physics and Astronomy, University of Oklahoma, Norman OK; United States of America.\\
% ...
$^{131}$LAL, Universit\'e Paris-Sud, CNRS/IN2P3, Universit\'e Paris-Saclay, Orsay; France.\\
% ...
$^{155}$Department of Physics and Astronomy, University of Sussex, Brighton; United Kingdom.\\
% ...
$^{160}$Raymond and Beverly Sackler School of Physics and Astronomy, Tel Aviv University, Tel Aviv; Israel.\\
% ...
$^{166}$Department of Physics, University of Toronto, Toronto ON; Canada.\\
% ...
$^{a}$ Also at Borough of Manhattan Community College, City University of New York, New York NY; United States of America.\\
$^{b}$ Also at Centre for High Performance Computing, CSIR Campus, Rosebank, Cape Town; South Africa.\\
$^{c}$ Also at CERN, Geneva; Switzerland.\\
$^{d}$ Also at CPPM, Aix-Marseille Universit\'e, CNRS/IN2P3, Marseille; France.\\
$^{e}$ Also at D\'epartement de Physique Nucl\'eaire et Corpusculaire, Universit\'e de Gen\`eve, Gen\`eve; Switzerland.\\
% ...
$^{*}$ Deceased

\end{flushleft}


\end{document}